\documentclass{article}

\usepackage{mathptmx}       % selects Times Roman as basic font
\usepackage{helvet}         % selects Helvetica as sans-serif font
\usepackage{courier}        % selects Courier as typewriter font
\usepackage{graphicx}        % standard LaTeX graphics tool
\usepackage[bottom]{footmisc}% places footnotes at page bottom
\usepackage[utf8]{inputenc}
\usepackage{amsmath}
\usepackage{amssymb}
\usepackage{cite}

\def\per{$\rm [\overline{1} \overline{1} 2]\,$}

\begin{document}
\begin{titlepage}
\begin{center}{\Large\bf Plasmons in one and two dimensions}\\[2ex]
{\large\bf H. Pfn\"ur$^*$\footnotemark[1], C. Tegenkamp$^*$, and L. Vattuone$^{**}$}\\[1ex]
\end{center}
{\small \em $^*$ Institut f\"ur Festk\"orperphysik, Abt. ATMOS, Leibniz Universit\"at 
Hannover, Appelstr. 2,\\ D-30167 Hannover, Germany\\
$^{**}$ Dipartimento di Fisica, Universit\`a di Genova, Via Dodecaneso 33, IT-16146, 
Genova, Italy}

\abstract{Plasmons in low-dimensional systems respresent an important tool for coupling energy 
into nanostructures and the localization of energy on the scale of only a few nanometers. Contrary 
to ordinary surface plasmons of metallic bulk materials, their dispersion goes to zero in the long 
wavelength limit, thus covering a broad range of energies from terahertz to near infrared, and from 
mesoscopic wavelengths down to just a few nanometers. Using specific and most characteristic 
examples, we review first the properties of plasmons in two-dimensional (2D) metallic layers from 
an experimental point of view. As demonstrated, tuning of their dispersion is possible by changes 
of charge carrier concentration in the partially filled 2D conduction bands, but for the 
relativistic electron gas like in graphene only in the long wavelength limit. For short 
wavelengths, on the other hand, the dispersion turns out to be independent of the position of the 
Fermi level with respect to the Dirac point. A linear dispersion, seen under the latter 
conditions in graphene, can also be obtained in non-relativistic electron gases by coupling between 
2D and 3D electronic systems. As a well investigated example, the acoustic surface plasmons in 
Shockley surface states, coupled with the bulk electronic system, are discussed. Also the 
introduction of anisotropy, e.g. by regular arrays of steps, seems to result in linearization (and 
to partial localization of the plasmons normal to the steps, depending on wavelengths). In 
quasi-one dimensional (1D) systems, such as arrays of gold chains on regularly stepped Si surfaces, 
only the dispersion is 1D, whereas shape and slope of the dispersion curves depend on the 2D 
distribution of charge within each terrace and on coupling between wires on different terraces. 
In other words, the form of the confining quasi-1D potential enters directly into the  
1D plasmon dispersion and gives new opportunities for tuning.}

\footnotetext[1]{email: pfnuer@fkp.uni-hannover.de, article to appear in Springer Handbook on Surface 
Science 2017}
\end{titlepage}

\section{Introduction}
\label{sec:1}
The long-range Coulomb interaction between valence electrons in metals results in collective plasma
oscillations, as pointed out already by Pines and Bohm \cite{Pines1952,Pines1956} as early as 1952. 
Plasmons exist in all dimensions \cite{Nagao2006}. The surface as the typical truncation of a bulk materials 
imposes a new boundary condition on these collective plasma oscillations resulting in localized states at the surface, called 
surface plasmons \cite{Ritchie1957,Raether1988}. In other words, the surface plasmon is the edge plasmon of a three-dimensional metal.  
Surface plasmons have a finite frequency in the long wavelength limit, which 
for a nearly free electron gas is located at $\omega_p/\sqrt{2}$ with $\omega_p$ the bulk plasmon frequency, and a small dispersion with
either positive or negative slope depending on the screening properties of the participating electronic orbitals \cite{Liebsch1997,Pitarke2007}. 
These properties of the surface plasmon also allow direct crossing of the surface plasmon dispersion with the light line 
and thus the direct interaction with light. The combined excitation of a surface plasmon in presence of electromagnetic radiation is 
called a surface plasmon-polariton (SPP), which represents the basis for the rapidly growing field of nanoplasmonics, opening the route for 
surface plasmon subwavelength optics \cite{Barnes2003}, surface plasmon enhanced spectroscopy \cite{Jahn2016} near field optics \cite{Kawata2001}
and chemical applications \cite{Grzelczak2008, Halas2011}.  

In this contribution, however, we will deal primarily with a different kind of collective low-energy excitations, generated in low-dimensional 
electron gases in one (1D) and two dimensions (2D). Most prominent examples are the 2D sheet plasmons of a metallic monolayer that were observed, e.g., 
in Ag or Dy monolayers on Si(111) \cite{Nagao2001, Rugeramigabo2008}, in monolayers of graphene on SiC or metal surfaces \cite{Politano2014}, 
but also the 2D plasmons of partially filled surface states found 
on the clean surfaces of Be(0001) \cite{Diaconescu2007,Jahn2012}, Cu(111) \cite{Pohl2010,Vattuone2012} 
and Au(111)\cite{Vattuone2013,Park2010}, which result in so-called acoustic surface plasmons (ASP). 
Their relevance has two aspects: It is worthwhile to study the specifics of plasmonic excitations in low-dimensional systems in more detail, 
since both partial localization and correlations are important \cite{Pitarke2007}. Therefore, 
a quantitative theory for plasmonic excitations going beyond the nearly-free  electron 
gas model (NFEG) \cite{Stern1967}  is highly desirable. Secondly, much higher in-plane confinements of excitations can be achieved with
plasmons associated with a 2D electron gas, since their dispersion is much flatter than that of SPP, which allows for extraordinary
confinement of energy from subterahertz to midinfrared
frequencies by use of appropriate nanostructures providing
the necessary momentum transfer.
The $\sqrt{q}$- dependence of dispersion, however, makes a distortionless
propagation of nonmonochromatic signals inherently impossible, since the different frequencies components
propagate at different velocities. This drawback can be overcome by use of systems that have a linear rather
than a square root dispersion. As we will illustrate below, a linear dependence of dispersion is always expected
for low-D plasmons that are coupled and shielded by other 2D or 3D electron gases \cite{Bill2003,Silkin2005}. 
This is, in particular, the case for metal surfaces supporting an electronic Shockley surface
state (SS) with band dispersion crossing the Fermi level \cite{Silkin2004}, which leads to the formation of the ASP .

The physics of experimentally accessible systems that exhibit quasi-1D properties are still largely unexplored.
The most obvious reason is their inherent instability, which allows their existence as isolated systems only at $T=0$K.
1D properties can, however,  be stabilized if embedded into a 2D or 3D environment. The interaction with these environments 
not only stabilizes these systems, their feedback on 1D properties opens a wide range of variability and allows to manipulate them.
Thus a variety of exotic phenomena such as charge and spin density waves becomes accessible and makes them quite interesting  
objects to be studied \cite{Gruner1994,Snijders2010}. 
In this context, the  arrays of atomic wires generated by self-assembly on semiconducting surfaces
\cite{Yeom1999,Crain2003,Crain2005,Tegenkamp2005, Blumenstein2011} are particularly attractive: they 
are not only  geometricallystrongly anisotropic \cite{Crain2003, Wippermann2010,Krawiec2010},
but they also exhibit quasi-1D Fermi surfaces and/or even 1D electronic transport properties under
certain conditions. Depending on their chemical interaction with the substrate, but also on the details of the
coupling of these wires mediated by the substrate, these ensembles reveal in part fundamentally different
characteristics. As an example, bundles of  In wires on Si(111) forming a $(4\times 1)$ reconstruction are
metallic at room temperature, but  undergo a metal-insulator transition to form  charge density waves below
130~K. In contrast, chains of   Au on Ge(100) seem to reveal under special conditions the  spectroscopic
the  signatures of a Luttinger liquid
without any  instabilities  at low temperature \cite{Blumenstein2011,Tanikawa2004}, but this behavior is still under discussion. 
Not only quasi 1D metal-insulator transitions 
are seen \cite{Tegenkamp2005, Tegenkamp2008}, but surface instabilities result in electronic stabilization of new facets. 
This stabilization is intimately coupled with changes in the electronic band structure leading, in some cases, to coupled spin 
and charge order  \cite{Tegenkamp2012,Brand2015}.

The 1D properties are not only visible by a combined view  at geometry and occupied electronic states,
they manifest themselves also in electronic excitations, which in the limit
of pure 1D behavior with its strong electronic correlations cannot be discriminated from collective plasmonic excitations \cite{Giamarchi2003}.
As we will show below, although the identification of purely 1D dispersion is easily possible 
\cite{Nagao2006,Nagao2007,Ruge2010,Block2011,Krieg2013}, the quantitative properties explicitly depend on the coupling to the 
environment. If the wires in arrays are essentially decoupled from each other, the strong confinement perpendicular to the chains 
leads to generation of electronic subbands. As a consequence, the simultaneous excitation of subbands  plasmons and intersubband 
plasmon  are observed \cite{Nagao2006,Ruge2010,Inaoka2005}. In fact, various forms of interwire coupling \cite{Block2011,Lichtenstein2016} 
lead partly to two-dimensional crossover in the quantitative dispersion properties.  

This chapter will give an overview on the properties of these low-D plasmons, discussion particularly characteristic examples. 
We will start with 2D sheet plasmons, concentrating on the plasmonic properties of the system most investigated in the recent past, graphene. 
Further emphasis will be given to low-D plasmons coupled to other electron gases, which leads to linearization in the form of ASP, but also
to crossover of dimensionality, depending on plasmonic wavelengths. Finally we turn to quasi-1D 
systems and their corresponding plasmons, and try to solve at the end the puzzle of broad loss 
peaks but still fairly large plasmonic lifetimes.

\section{Sheet plasmons}
\label{sec:2}
Historically, sheet plasmons were first observed as standing waves for a dilute 2D electron system (2DES) on a liquid
He surface \cite{Grimes1976} at extremely low electron densities (few $10^6$/cm$^2$). The square root dependence of the 2D plasmon frequency was 
already seen in inversion layers of Si-MOS field effect transistors and in GaAlAs quantum well structures
\cite{Allen1977,Jusserand1990}. These artificially tailored systems allowed only large electron
spacing or long Fermi wavelengths ($\ge 1000$\AA). Their energy dispersions were studied in a tiny wave number region
$q_\parallel \le 0.01$\AA$^{-1}$ where they can be  described by classical local response theory and are also free from the lifetime
broadening caused by Landau damping. Plasmons of Shockley surface states are other examples apart from 2D metallic layers or stacks. All these systems 
have in common that they are not free standing, but are either on top or embedded into a carrier material. Depending on its conductive 
properties,  this material is responsible for shielding and modifications of the electric field caused by the plasmons. 
Therefore, before we discuss specific examples, let us briefly resume the theoretical ideas explained in more detail 
in the chapter by V.M. Silkin. 

\subsection{Plasmon dispersion for the 2D electron gas, influence by the environment}
For a purely 2D nearly-free electron gas (NFEG), i.e. when no bulk electrons are present, theory predicts the existence of a surface plasmon exhibiting 
in the limit $q_\parallel \rightarrow 0$ a square root dispersion. {\textit{$q_{\parallel}$}} is the parallel component of the wavevector of the excitation created in a 
scattering event. At least the next order, however, will be important in some cases (see below). 
Following Stern \cite{Stern1967}, 
%in the absence of the 3D substrate, a Shockley surface state would support a 2D collective oscillation, 
the energy of the corresponding plasmon being given by:
\begin{equation}
\omega^2_{2D} = e^2 \frac{E_F}{2\epsilon' \pi\hbar^2} q_\parallel + \frac{3}{4} v_F^2 q_\parallel^2 + ... \label{eq1}
\end{equation}
with the Fermi energy $E_F$. While for a free standing 2D system $\epsilon'$ is just equal to the vacuum dielectric constant $\epsilon_0$;  when the 2D electron
gas is embedded in an insulating environment with dielectric function $\epsilon$, then  $\epsilon'= \epsilon_0 \cdot \epsilon$. 

The first term in eq. \ref{eq1} is identical to the result for a thin metallic film obtained by the classical response theory \cite{Ritchie1957}, whereas the second term (and higher order terms) 
result from the correlations in a 2D Fermi gas and corresponding changes in the excitation spectrum. 
For the free electron gas, the 2D electron density of occupied electronic states 
is related to $E_F$ and to the effective mass $m^\star$ by  
\begin{equation}
n_{2D}=\frac{E_F m^\star}{\pi \hbar^2} 
\end{equation}
For a relativistic electron gas with linear dispersion ($E = \hbar v_F q_\parallel$), $n_{2D}$ depends , on the contrary, quadratically on $E_F$: 
\begin{equation}
n_{2D} = \frac{E_F^2}{\pi \hbar^2 v_F^2}
\end{equation}
so that the leading term in eq.\,\ref{eq1} of the plasmon frequency at small $q_\parallel$ depends on electron density as $n_{2D}^{1/4}$. 
Interestingly, the second term of eq.\,\ref{eq1} does not depend on the electron density in this case, since $v_F$ is constant. 
As we shall show belowe, experiments confirm this prediction.  

Screening by the environment due to embedding into or adsorption of a 2D layer on a carrier material, or due to the presence of a bulk in case of a surface state 
will lead to modifications of eq.\,\ref{eq1}. In the simplest case of adsorption of the 2D layer on an insulating material the   resonances are far
away from the plasmon frequencies and screening effects can be described by a dielectric constant, which is taken as the average of 
the dielectric constant of this material and vacuum, so that $\epsilon'$ now reads: $\epsilon' = \epsilon_0(\epsilon + 1)/2$.  

In case of a metallic surface state screened by the presence of 3D electrons, theory predicts in the long wavelength limit, the existence of two plasmons \cite{Pitarke2007}: 
One of them corresponds to the high-frequency oscillation $\omega_h$,
in which 2D and 3D electrons oscillate in phase with each other. 
The other one corresponds to a low-frequency acoustic oscillation in which both 2D and 3D electrons oscillate out of
phase, characterized by a linear dispersion:
\begin{equation}
\omega_l = \alpha v_F^{2D} q_\parallel
\label{disp_asp}
\end{equation}
where $\alpha$ is a constant with a value close to 1 and $v_F$ is the 2D Fermi velocity.
As we shall show below, in the most common case, i.e. for Be(0001) \cite{Diaconescu2007}, the 2D Fermi velocity 
will be lower than the 3D one so that $\alpha$ is slightly larger than 1.
However, there are cases, e.g. Au(111) \cite{Vattuone2013} and most likely also Cu(111) \cite{Pischel2013}, for which 
the 2D Fermi velocity is higher than in 3D so that $\alpha$ is slightly lower than 1.

\begin{figure}[htb]
\centering
\includegraphics[width=0.8\linewidth]{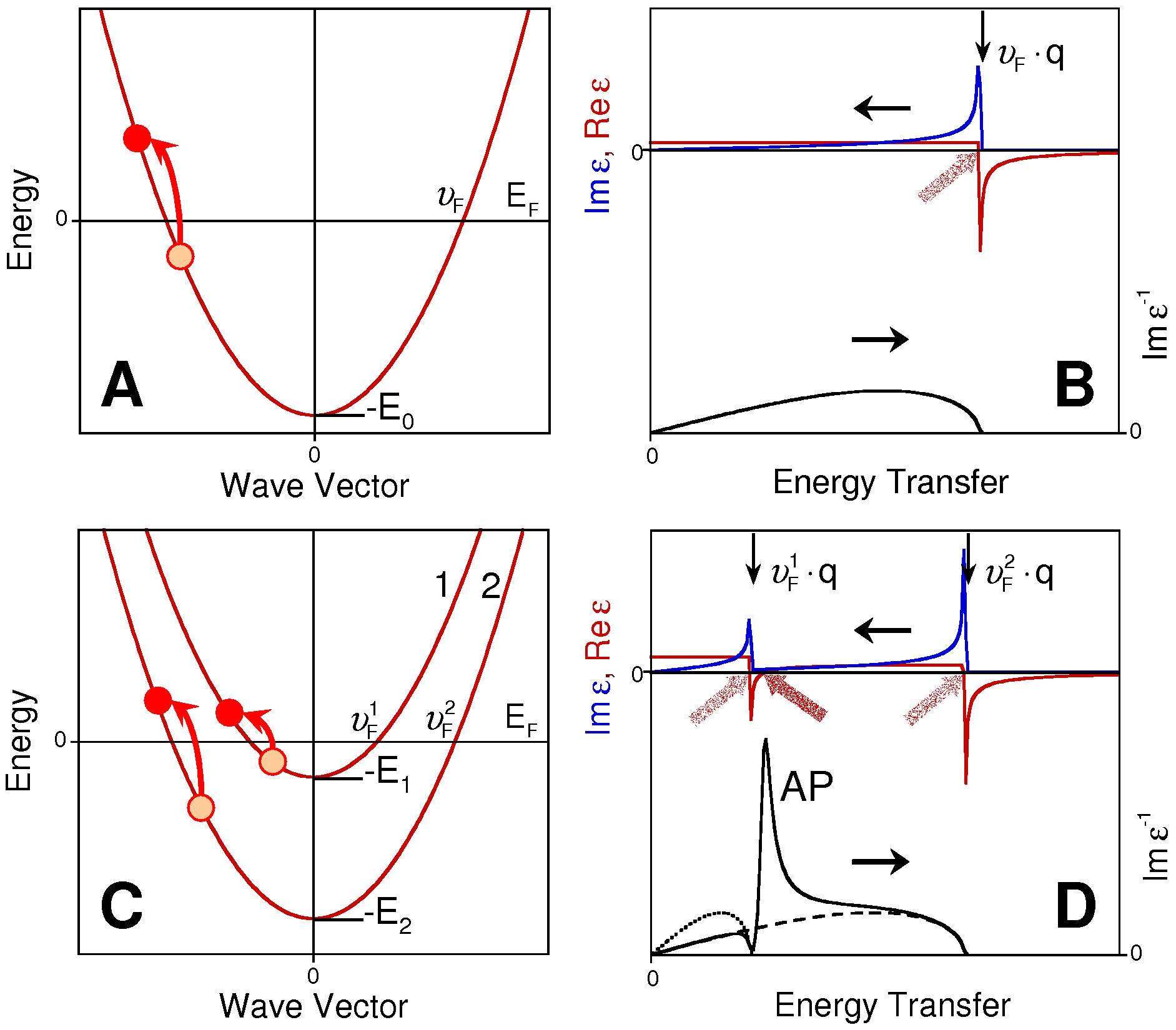}
\caption[]{Schematic of electronic excitations and decay in single- (a,b) and double-component (c,d) 2D electron 
gases: Energy dispersion as a function of wave vector (panels a
and c) with single electron intra-band transitions (red arrows). In
(b) and (d), the dielectric function $\epsilon$ with real (red lines)
and imaginary parts (blue lines), and the loss function
$Im(\frac{1}{\epsilon})$ (black lines) are plotted for an arbitrary, but given
small momentum transfer. Unobservable low-energy resonances are indicated by light red arrows. The observable 
feature, characterised by Re$(\epsilon) = 0$ and a maximum in the loss function (and hence a small value in $Im(\epsilon)$) 
is indicated by a dark red arrow.
Reprinted with permission from ref. \cite{Vattuone2013}}
\label{Fig_1and2egases} 
% fig 1 \cite{Vattuone2013}
\end{figure}
In order to understand the basic mechanism behind the ASP, let us compare the behavior of a conventional 2D electron gas 
with a parabolic dispersion (see Fig.\,\ref{Fig_1and2egases} a) with an electron gas consisting of two components 
with different Fermi velocities (see Fig.\ref{Fig_1and2egases} c). The corresponding dielectric function
$\epsilon(\omega,\textbf{k})$ for a given wavevector \textbf{k} (in the long wavelength limit) and the loss 
function $Im(1/\epsilon)$ are shown in the right panels of the same figure.
For the 2D electron gas with only one component the low energy zero crossing of $Re [\epsilon(\omega,\textbf{k})]$ 
is not associated with a peak in the loss function and thus corresponds to the excitation of incoherent intra-band electron hole pairs. 
For a two-component electron gas there exist three zero crossings of $Re [\epsilon(\omega,\textbf{k})]$ at low energy.
One of them is associated with a  peak marked in the loss function $Im(1/\epsilon)$.
Since it lies just above the edge of single particle transitions within the band with the lower Fermi velocity, it
corresponds to a collective mode screened by the electrons
located in the other band. In the long wavelength limit, i.e. for $q_\parallel \rightarrow 0$, the energies of both peaks 
in $Im(\epsilon)$ disperse linearly with q. The peak in the loss function, being limited by such linearly dispersing 
peaks, exhibits also a linear dispersion and satisfies thus eq.\,\ref{disp_asp}.
In the case of a Shockley surface state at a metal surface, the model still captures the essential physics, provided 
that one of the two components of the electron gas is associated to the 3D electron system and that the coexistence 
of 3D and 2D electron gases in the same volume is properly considered. 

It is worth mentioning here that this scenario of interacting electron gases is not limited to the specific case just discussed. E.g., stacking of 
identical layers, separated by insulating spacers, considered theoretically in ref.\,\cite{Bill2003} also leads to coupling between electron gases and 
to linearization of plasmon dispersion. 

The plasmon dispersion can be measured experimentally by using (highly resolved) Electron Energy Loss Spectroscopy (HREELS).
An instrument with simultaneous high energy and high momentum resolution is called ELS-LEED \cite{Claus1992}. 
Electrons with energy $E_i$ impinge at an angle $\theta_i$ with respect to the surface normal. Electron can scatter elastically or inelastically. 
The energy spectrum of the electrons scattered at an angle $\theta_s$ from the surface normal can contain peaks at energy $E_s$. 
The energy lost (or gained) corresponds to the creation (or annihilation) of a collective excitation of energy $E$ such that:
\begin{equation}
E_i = E+E_s
\end{equation}
Note that $E$ is positive when the excitation is created and negative if it is destroyed in the scattering event.
%\boldmath{\textit{$\mu_{el}$}}\unboldmath and \textbf{\textit{E}}. The values of %$\mu_{el}$ and E stand for $|\boldmath{\textit{$\mu_{el}$}}\unboldmath|$ and $|%\textbf{\textit{E}}|$, respectively.
The conservation of momentum in direction parallel to the surface reads:
\begin{equation}
\boldmath{k_{i,\parallel}}=k_{s,\parallel}+q_{\parallel}+g_{\parallel}
\label{eq_parmomentum}
\end{equation}
\unboldmath
where 
\boldmath{\textit{$g_{\parallel}$}}\unboldmath\ is any reciprocal lattice vector of the surface unit cell.
Since before and after scattering the electron is a free particle (the same relationship holding also for any free particle with 
non-vanishing mass in the non relativistic limit) the modulus $k_{i,\parallel}$ of the parallel component \boldmath{\textit{$q_{\parallel}$}}\unboldmath of the wavevector is given by:
\begin{equation}
k_{i,\parallel}=\frac {\sqrt{2 m E_i}}{\hbar} cos \theta_i
\end{equation}
The modulus $k_{s,\parallel}$ is given by a similar relationship. 
It is thus possible to obtain $\boldmath{q_{\parallel}}\unboldmath$ and $E$, i.e. the dispersion relation of the collective excitation 
(or the energy and the momentum of the particle) created in the scattering event.
Experiments performed for Ag on Si \cite{Nagao2001} and for $\rm DySi_2$ \cite{Ruge2008} confirmed the prediction.
At very long wavelengths, 2D plasmons have low energy. However they are unable to affect electron-phonon interaction 
and phonon dynamics close to the Fermi energy because of the square-root dispersion.

\subsection{Sheet plasmons in 2D}
\subsubsection{Metallic layers}
\begin{figure}[tb]
\centering
{\bf a)}\includegraphics[width=0.4\linewidth]{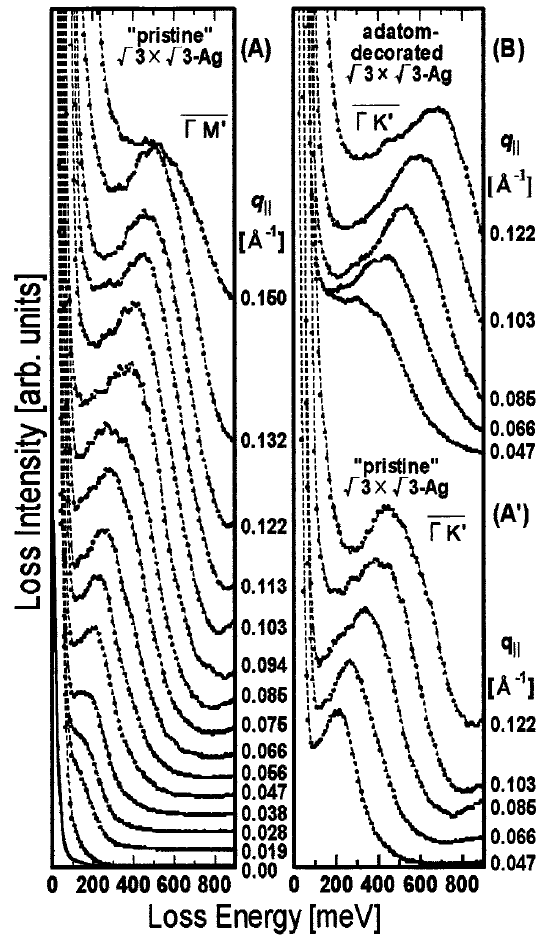}\hspace{0.7cm}
{\bf b)}\includegraphics[width=0.44\linewidth]{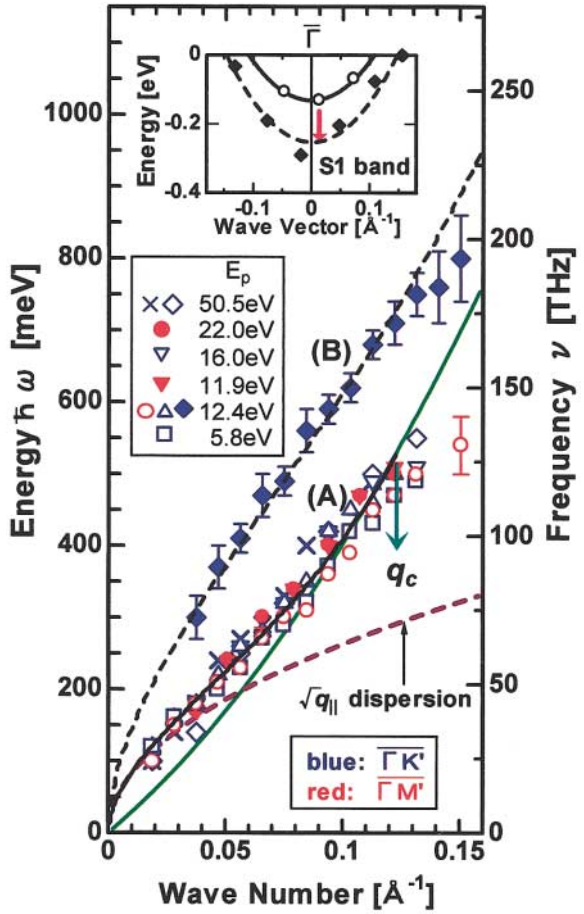}
\caption{\label{agsi} a) EELS spectra taken at an incident electron energy of $E_p = 12.4$\,eV 
(from ref.\cite{Nagao2001} with permission). For better visibility the curves were shifted, the momentum transfer is given as parameter.  
(A) Scans along the $\overline{\Gamma M}$ direction (A') and (B): Blueshift
of losses due to the additional adatom decoration. 
b) Plasmon dispersions of (A) a $\sqrt{3}\times \sqrt{3}$R30$^\circ$ ordered surface at a coverage slightly above 1 ML and 
(B) after further decoration with 0.15 ML of additional Ag. The bold solid and dashed curves are the fits within the nearly
free electron gas model in RPA approximation. Green curve: upper edge of the single-particle excitation
continuum.  Inset: Schematic of the S1 surface band that is shifted below $E_F$ by Ag coverages exceeding 1ML.}
\end{figure} 

Here we discuss the prototype system Ag/Si(111) \cite{Nagao2001}. Similar results have been obtained for Dy/Si(111) \cite{Ruge2008}.
The flat Ag/Si(111) system has been a prototype for the investigations of surface conductance on 
a low-doped Si substrate \cite{Hasegawa2000}, in which metallic conductance has been shown to be associated with the 
formation of a $\sqrt{3}\times\sqrt{3}$R$30^\circ$ structure. Conductance can be changed by adding further Ag atoms. 
In fact, such $\sqrt{3}$-structure  turned out to be a semi-metal \cite{Crain2005}, and metalliticy is attained only by adding  surplus Ag atoms 
 to this structure  \cite{Hasegawa2000}. The most natural explanation for this self-doping phenomenon 
seems to be the formation of a lattice gas in the second layer, as also suggested by a HREELS study as  a function of Ag 
concentration \cite{Liu2009}. The investigations on stepped Si(557) yield  however a somewhat different picture (see below).

In any case, 2D metallicity can be generated in this system. A direct proof is the existence of a plasmonic excitation whose energy 
goes to zero at long wavelengths, no further excitation being necessary to create this plasmon. The first test for such a purely 2D system 
has indeed been carried out in Ag/Si(111) at monolayer concentration \cite{Nagao2001}. The main results are shown in Fig.\,\ref{agsi}. 

As seen in Fig.\,\ref{agsi}b), the dispersion can be well described by a 2D non-relativistic NFEG using the two terms given explictly in eq.\,\ref{eq1}. 
Within errors there was no angular dependence of the dispersion, but the slope of dispersion changes when further Ag was added to the 
(already self-doped)  $\sqrt{3}\times \sqrt{3}$R30$^\circ$-Ag structure. From their fits the authors determined electron concentrations 
between 1.9 and $8\times 10^{13}$\,cm$^{-2}$ and an effective electron mass around $0.3 m_e$. This result is in good agreement with 
data from angular-resolved photoemission (ARPES) for the same system \cite{Crain2005}. From these numbers an effective electron transfer of about 
1/3 of an electron per surplus Ag atom was deduced. For the present case, the screening by the substrate can thus be 
accurately described by an effective dielectric constant of the interface.  

\subsubsection{Plasmons in the 2D relativistic electron gas}
Since its final discovery in 2004 \cite{Castro2009} graphene has become one of the most studied materials in nanoscience. The 
coupling of light into flat and nanostructured graphene in the terahertz and far infrared frequency regime, using the 
plasmonic properties of graphene in order to form surface plasmon polaritons (SPP) has been a widely
covered subject \cite{Wang2011,Abajo2014,Emani2015}, and many potential applications have been suggested 
\cite{Achanta2015,Chou2014,Eng2015,Gao2015}. We refer the reader to these excellent reviews and will not elaborate on these topics here, since 
although in this range of frequencies sub-wavelength phenomena (compared with optical and infrared wavelengths) can be observed, 
the typical wavelengths of several $\mu$m do not allow to enter the nanoscale of 10 nm and below.

However, there is a unique property of 2D plasmons that have became obvious by studying 2D plasmons in graphene. Here we concentrate again on the 
low-energy plasmons. Other plasmons in graphene, in particular the standard $\pi$ plasmon excitations has already been reviewed before \cite{Politano2014} 
and we refer the reader to this literature. At low $q_\parallel$ (typically in the range of $\mu$m$^{-1}$), i.e. at plasmon frequencies in the terahertz range, 
a series of studies have exploited the fact that the plasmon frequencies of plasmon polaritons, i.e. the combined excitation of 
a plasmon in an external electromagnetic field,  can be tuned by changing the carrier concentration in the graphene layer. 
This can most conveniently be done by using a gate voltage \cite{Fei2012,Woessner2015}. Thus plasmon polaritons at a wavelength of a 
few hundred nanometers can be excited with light in the far to middle infrared range. This result is 
attractive because its  relatively low damping:  decay lengths reaching  1$\mu$m allow indeed to 
foresee many perspectives potential applications such as tunable infrared lasers \cite{Pollini2016}, 
plasmonic quasicrystals \cite{Achanta2015}, 
ultrasensitive detection down to the single-molecule level \cite{Kneipp1997}, improved photovoltaics,\cite{Atwater2010}, or nonlinear optics \cite{Harutyunyan2010}.
This tunability, however, is subject to certain physical limitations that we will describe further below. 

\begin{figure}
\includegraphics[width=0.45\linewidth]{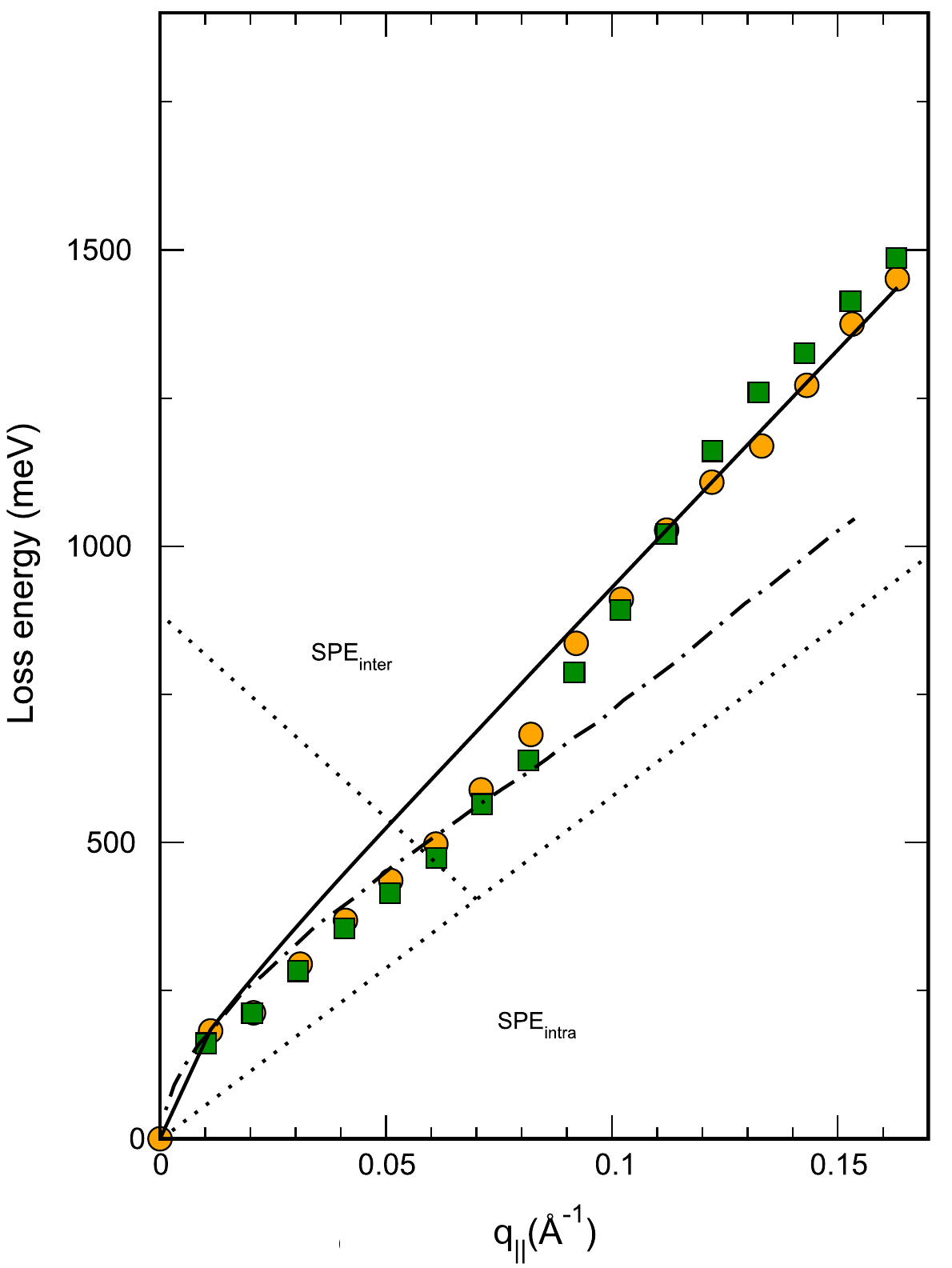}
\includegraphics[width=0.45\linewidth]{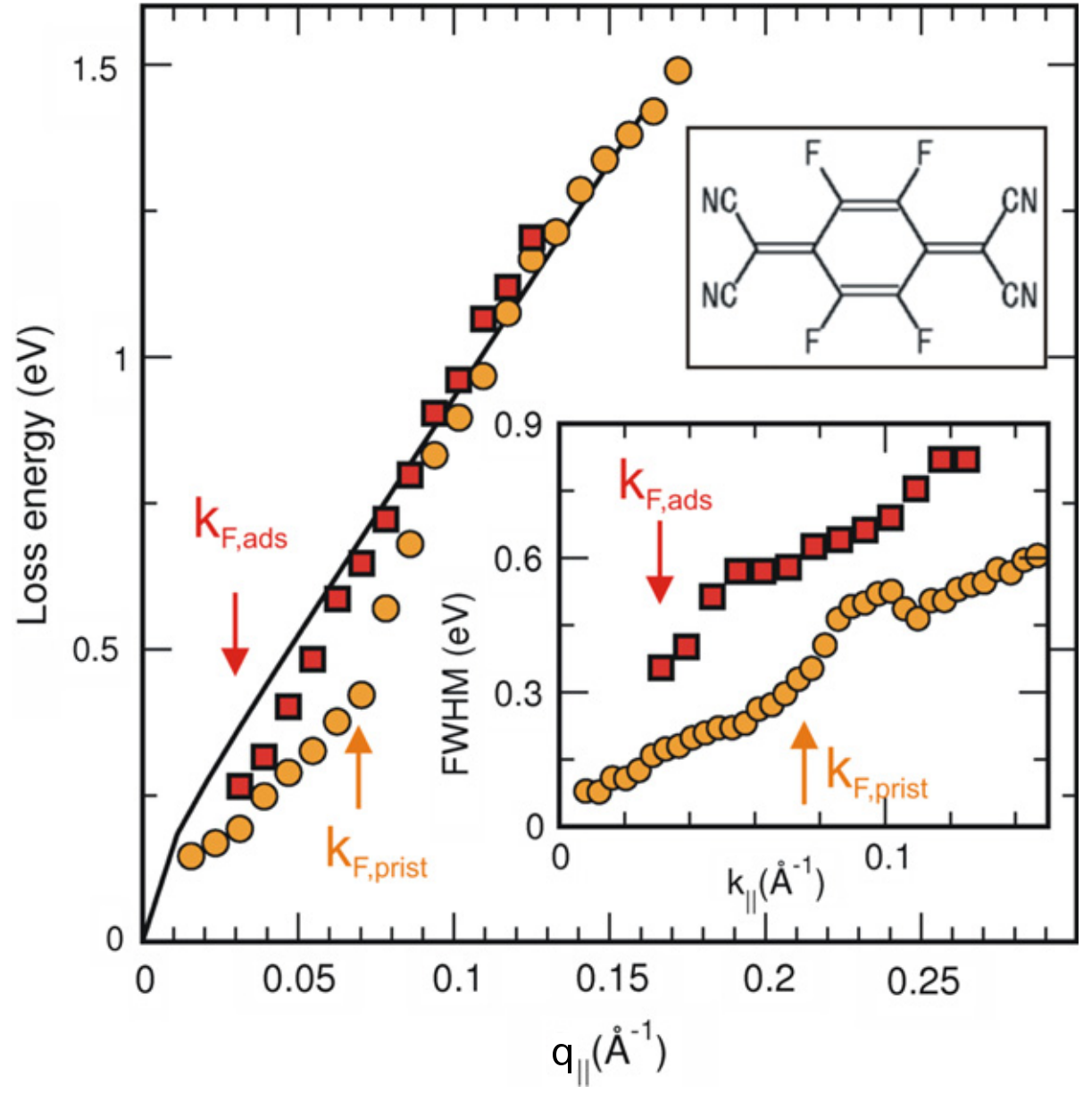}
\caption{\label{gdop} Left: Plasmon dispersion for 1 ML graphene layer grown on H-etched SiC samples, measured at
300\,K ($\square$) and at 80\,K ($\bigcirc$).
Right: Graphene plasmon dispersion before ($\bigcirc$) and after ($\square$) F4-TCNQ doping. 
Lower inset: Changes of k-dependent halfwidths (same symbols as in main figure). Upper inset: stereographic model of F4-TCNQ.
}
\end{figure}

Surprisingly, this tunability seems to completely disappear, when $q_\parallel$ is in the range of 1 nm$^{-1}$. 
An example is shown for graphene on Si-terminated 6H-SiC(0001) in Fig.\,\ref{gdop}. The left part shows the plasmon dispersion, recorded 
with  electron energy loss spectroscopy with simultaneous high momentum and high energy resolution \cite{Tegenkamp2011}. These results agree well with 
those of ref.\,\cite{Liu2008}.
The first layer of graphene on SiC grows on a carbon buffer layer. It turns out to be n-doped with a doping 
concentration of $1\times 10^{13}$\,cm$^{-2}$, and the Fermi level under 
these conditions is about 400 meV above the Dirac point \cite{Tegenkamp2011}. 

At very low $q_\parallel$ the dispersion is compatible with a $\sqrt{q_\parallel}$ dispersion,  as predicted by the first term of eq.\,\ref{eq1}. 
It clearly deviates from this function already at $q_\parallel < 0.03$\,\AA$^{-1}$, similar to the Ag/Si(111) case discussed above. 
For  $q_\parallel > 0.1$\,\AA$^{-1}$, on the other hand,
there is a linear slope of $1.35 \pm 0.1 \times 10^6$ m/s, very close to the Fermi velocity in graphene on SiC \cite{Hwang2012}. 
Between these two regimes there is a characteristic cusp that was ascribed to resonant damping since the plasmon dispersion 
enters the continuum of single particle interband transitions.

\begin{figure}[tb]
\centering
\includegraphics[width=0.5\linewidth]{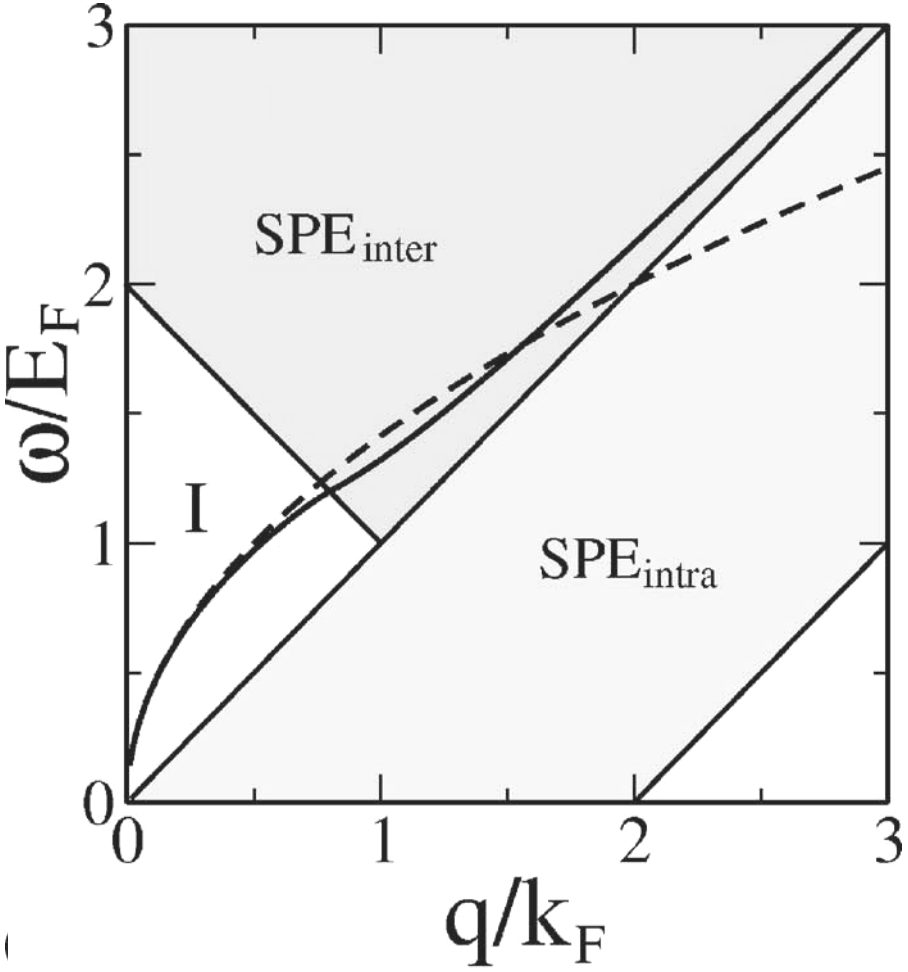}
\caption{\label{disp-theory} Calculated plasmon dispersion for monolayer graphene within RPA on a substrate characterized by 
$\epsilon = 2.5$ (from ref.\,\cite{Hwang2007}, with permission). The dashed line corresponds to the plasmon dispersion of a 2D free 
electron gas with an electron density of $1\times 10^{12}$\,cm$^{-2}$. Shaded areas correspond to the continuum of 
intra- and interband single particle excitations.}
\end{figure}

Qualitatively, the experimental dispersion curve follows the theoretical curve developed for a 2D Fermi gas 
of relativistic electrons within the random phase approximation (RPA) \cite{Hwang2007}. 
The resonant damping has not been considered in theory.  
Nevertheless, there is still a cusp seen in theory (see Fig.\,\ref{disp-theory}), which is a characteristic property 
of the relativistic electron gas. It is much less pronounced than in experiment because of the neglect of damping in the theoretical calculation. 
Therefore, quantitative agreement between theory and experiment cannot be expected.   

The main qualitative difference between the non-relativistic and the relativistic 2D electron gas comes from the second term in the expansion 
of plasmon dispersion, which becomes dominant for large $q_\parallel$. This term disperses linearly with a slope $\propto v_F$, 
as already outlined above (see eq.\ref{eq1}). 
For the relativistic electron gas of graphene, this limit is reached for $q_\parallel / k_F > 1$ \cite{Hwang2007}. 
If this second term dominates, its slope, according to eq.\ref{eq1}, does not depend on the embedding environment. 
Even more important, for the relativistic electron gas with a constant Fermi velocity this term 
is independent of the doping level.  Thus the slope of the dispersion in this limit should be usable as a calibration standard.
This prediction can easily be tested. 

One test is shown in the right part of Fig.\,\ref{gdop}. Here the plasmon dispersion curves before and 
after covering the 
graphene layer with about one percent of a monolayer of 4F-TCNQ molecules  at 300 K are compared. 
F4-TCNQ is known to reduce the carrier density of graphene and hence to shift $E_F$ downwards \cite{Chen2007}. 
As expected, the dip shifts to lower $q_\parallel$ values (to $\approx 0.03$\AA$^{-1}$). This shift is
also seen in the resonance of the FWHM. Since the added molecules act as scattering centers, the loss structure is further
broadened. This reduction of the doping concentration to $\approx 3\times 10^{12}$cm$^{-2}$ also reduces 
both $E_F$ and $k_F$ correspondingly. Therefore, the transition to the linear part of the dispersion happens at smaller 
$q_\parallel$, as expected from the considerations mentioned above, and the linear range is extended. 
The average slope of the linear part, however, remains unchanged, in accordance with a constant Fermi velocity and eq.\,\ref{eq1}.
A very similar behavior has been found for a quasi free standing graphene layer on SiC(0001), generated with a graphene buffer
layer by intercalation with hydrogen, by gradual desorption of the hydrogen \cite{Langer2012}, which mainly shifts the Fermi level
and thus the doping concentration.

\begin{figure}[tb]
\centering
\includegraphics[width=1\linewidth]{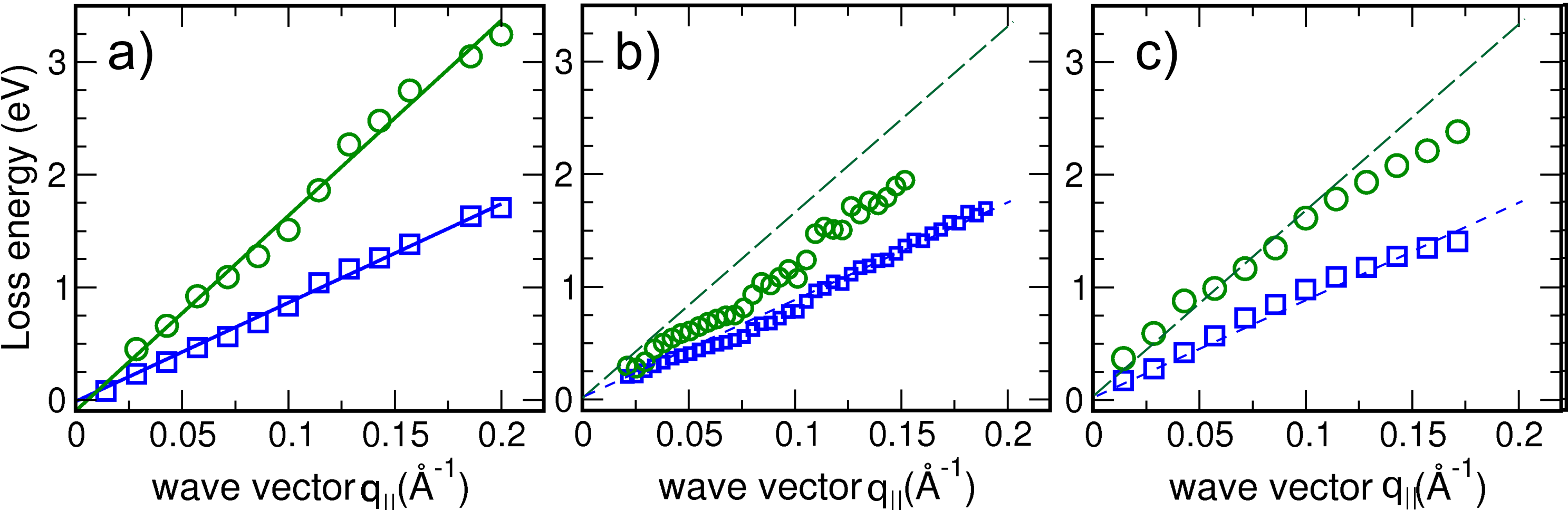}
\caption{\label{comp} a) Plasmon dispersion of the two plasmonic branches of graphene on Ir(111). 
b) same type of data evaluation for graphene on SiC(0001) (n-doped layer $n \approx 10^{13}$ cm$^{-2}$)
c) Results for Na intercalated graphene on Ir(111). The average linear slopes of a) are drawn as
dashed lines in the other graphs. While the lower branches 
correspond to the ``standard'' (or monopole) sheet plasmon, the others marked by green symbols are multipole plasmons.}
\end{figure}

A second example of this finding is shown by comparing these results with those obtained for graphene on Ir(111) \cite{Langer2011,Pfnuer2011}.
The graphene layer on Ir grows self-limiting as one layer by decomposition of hydrocarbons (ethylene) at temperatures around 1320
K, at a pressure of $2\times 10^{-7}$\,Pa. It is characterized by a Fermi level very close to the Dirac point \cite{Pletikosic2009,Rusponi2010}, i.e. a doping
concentration of less than $1\times 10^{11}$\,cm$^{-2}$. Therefore, and in accordance with the results just presented, only a linear slope 
can be observed within the $q_\parallel$ range resolvable with ELSLEED. This result is shown in Fig.\,\ref{comp} together with the results obtained 
after doping the layer by intercalation with half a monolayer of Na and in comparison with those from SiC. Here we concentrate first
on the lower dispersion branch. 
When comparing the slopes of the quasi linear dispersion for graphene on Ir(111) with that on SiC(0001) (middle panel of Fig.\,\ref{comp}), they 
are virtually identical. While the more effective shielding of the metallic surface compared to SiC should lower the slope of dispersion \cite{Hwang2012},
this shielding is not static and leads to a more complicated interaction between the 2D electron gas of graphene and  
the 3D electron gas in the conduction bands of Ir, which may compensate the expected red shift. 
However, this effect has not been studied quantitatively in this system. Completely linear dispersion has also been found for graphene on Pt(111) 
\cite{Politano2011,Politano2014}. The slope, however, is about 15\% smaller than in the examples just shown, possibly due to different interaction 
and shielding compared with the Ir(111) substrate.

Intercalation of graphene on Ir(111) by half a monolayer of Na has the effect of strong p-doping of graphene due a shift of the Fermi level of graphene 
by nearly 1\,eV. As a consequence, linearity of plasmon dispersion is only seen over a reduced $q_\parallel$ range upto 0.1\,\AA$^{-1}$, whereas for 
higher $q_\parallel$ values there is a tendency for levelling off. The slope measured in the linear $q_\parallel$ range is about 10\% 
larger compared to graphene on clean Ir(111). This is a remarkably small effect considering the change of doping level by about four orders of magnitude.  
On the other hand, both findings indicate that Na intercalation not only rigidly shifts the graphene bands, but Na interacts more strongly both 
with graphene and Ir causing the deviations measured. The increase in measured halfwidths of plasmon losses also point in this direction \cite{Pfnuer2011}.

In view of eq.\,\ref{eq1}, these results are not expected, since the doping level  considerably increases the first term of this equation,  so that it should dominate over a larger range of $q_\parallel$. 
A clear dependence of the plasmon dispersion has indeed been seen on SiC(0001) by varying 
amounts of intercalated potassium \cite{Shin2011}. In fact, due to the high doping level studied there, this first term of eq.\,\ref{eq1} dominates most of 
those results, and there is semi-quantitative agreement with the predictions of eq.\,\ref{eq1} with this assumption. 
At this point, we want to stress, however, that the findings of ref.\cite{Shin2011} are not in contradiction to the results 
found on Ir and the others discussed above. On the contrary, they fully agree with the predictions of eq.\,\ref{eq1}.
Since on a metal the screening is much more effective, the dielectric function is up to 2 orders of magnitude larger than for SiC, so that  the first term of eq.\,\ref{eq1} is smaller by the same factor
on metals like Ir or Pt compared to SiC:  the linear term in $q_\parallel$ thus dominates already at small $q_\parallel$. 
Moreover the interaction with the shielding metal electrons leads to further linearization of the dispersion curves \cite{Bill2003,Silkin2005},  a mechanism not present on a semiconductor like SiC. 

From these results we conclude that the tunability of the plasmon frequencies in the relativistic 2D 
electron gas is only possible  if the  term $\propto \sqrt{q_\parallel}$ dominates  the dispersion. 
The strong sensitivity of this term to the environment via its dielectric function $\epsilon$ not 
only influences the accessible frequencies, but also the range of tunablity. This range can easily 
be exceeded particularly on metals, where
the term linear in $q_\parallel$ is responsible for the insensitivity of plasmon dispersion to the doping 
level and to the environment. This robustness hasn't been exploited yet, but is quite important for many potential applications.

2D plasmons were also shown to couple to other excitations, in particular to phonons and single electron excitations. 
The enhanced coupling of plasmons to single particle excitations is quite well known: similarly to the situation described above \cite{Tegenkamp2011} due to 
opening of new decay channels, it causes resonantly enhanced damping. Similar phenomena have also been observed in photoemission, but here 
the linear dispersion of (ideally) non-interacting single quasi-electrons and -holes is modified 
close to the Dirac point by the interaction with the plasmon due to excitations of so-called 
plasmarons \cite{Bostwick2010,Walter2011}. 
F
or the coupling between sheet plasmons and phonons a prominent example is the coupling in graphene to the surface optical phonon of the 
underlying SiC on the Si-terminated surface \cite{Koch2010, Koch2016}. 
The coupling of this phonon turned out to be remarkably robust. It cannot be quenched by intercalation and modifications of the intercalation.
Instead, new coupled modes appear that are due to the coupling between the sheet plasmon with phonons of the intercalated material. 
Only for bilayer graphene coupling of the plasmon with the internal LO phonon was found to be detectable. These coupling may give a further handle to tailoring plasmonic properties. 

Finally, we turn to the observation of multipole modes, which were identified for graphene on Ir(111) and on SiC(0001) \cite{Pfnuer2011}. As seen from
Fig.\,\ref{comp}, these modes follow closely the dispersion of the basic mode, but with a different slope depending on the substrate material. On the 
metal the ratio between slopes is roughly 2, while on SiC it is reduced to 1.4. The halfwidths of the multipole losses turned out to be 
essentially the same as for the basic mode. Both modes appear at the same $q_\parallel$ so that it seems that both modes are excited 
by the same mechanism. It is indeed only a matter of probability 
which one dominates. The different ratios of slope found on the metal and on the semiconductor may, on the other hand, be due again to different efficiency of screening. 
These multipole modes in low-dimensional systems have not been explicitly considered by theory yet, even if this improvement is highly desirable in order to get deeper insight. 
%%%%%%%%%%%%%%%%%%%%%%%%%%%%%%%%%%%%%

\subsubsection{The acoustic surface plasmon (ASP) for Be(0001), Cu(111) and Au(111)}
\begin{figure}[htb]
\centering
\includegraphics[width=0.6\linewidth]{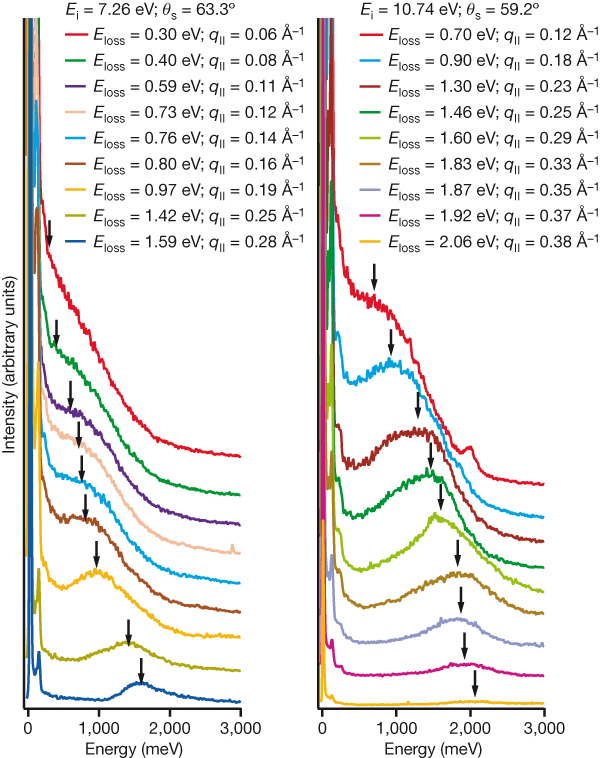}
\caption[]{HREEL spectra recorded for different scattering conditions starting from the specular direction 
(uppermost spectra) and moving out of specular for Be(0001). The arrow indicate the position of the energy loss feature.
Reprinted with permission from ref. \cite{Diaconescu2007}.
\label{Fig_Be(0001)_spectra}} 
% figg 1 \cite{Diaconescu2007}
\end{figure}

\begin{figure}[tb]
\centering
\includegraphics[width=1\linewidth]{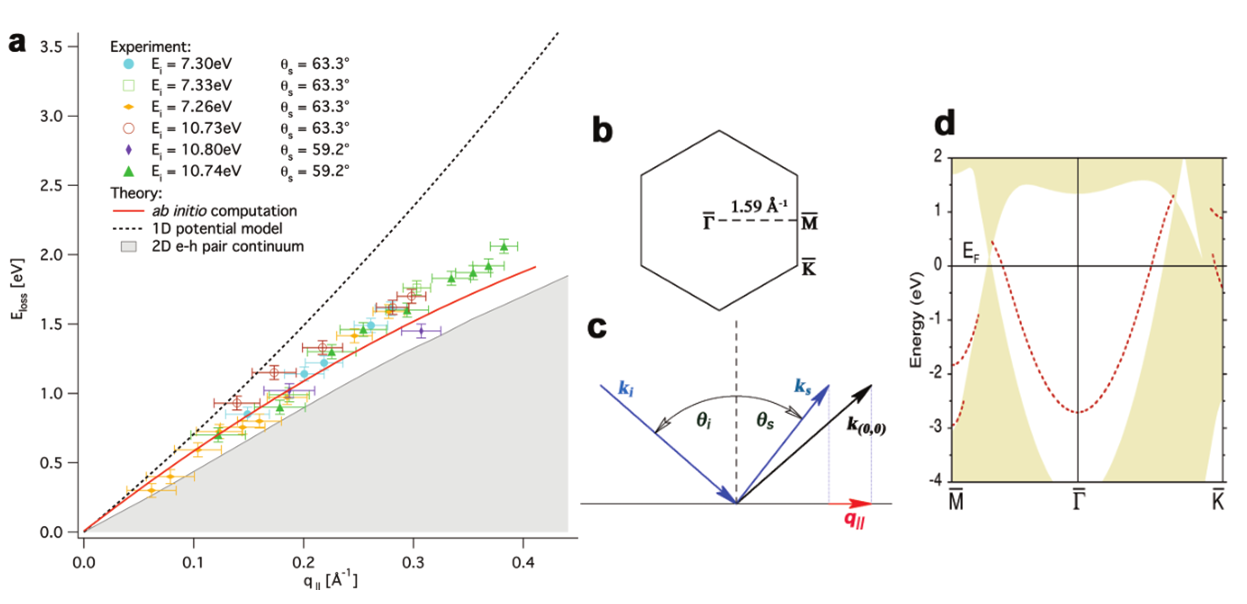}
\caption{Experimentally measured dispersion curve showing the energy of the ASP vs. parallel wavevector. The 
theoretical predictions are also shown: dotted black line indicates the simpler 1D model while the continuous 
red line indicate the result of a more accurate {\it ab initio} calculation. The shaded area indicates the 
continuum of electron - hole excitations; b) 2D Brillouin zone; c) schematic of the geometry of the experiment; 
d) Electronic structure of Be(0001) with the surface Shockley state in the gap.
Reprinted with permission from ref. \cite{Diaconescu2007}.
\label{Fig_Be(0001)}}
% figg 2 \cite{Diaconescu2007}
\end{figure}

As mentioned in the Introduction, a new mechanism of linearization of plasmon dispersion comes into play when excitations of 
different electron gases can be coupled. An important example is the 2D electron gas of a Shockley surface state, which exists, e.g.,
on the close-packed surfaces of the noble metals (Ag(111), Au(111) and Cu(111)), but also on Be(0001). The 2D sheet plasmon 
will inevitably be screened by the underlying 3D electron gas of the metal. 
{}
The ASP was first observed experimentally on Be(0001) with HREELS. 
Experimental data are shown in Fig. \ref{Fig_Be(0001)_spectra} along the $\overline{\Gamma M}$ direction for two different 
kinetic energies (7.26 and 10.74 eV) while keeping the scattering angle $\theta_s$  fixed 
(63.3$^{\circ}$ and 59.2 $^{\circ}$) and  changing the angle of incidence $\theta_i$. The parallel 
momentum is obtained using eq. \ref{eq_parmomentum}.
It is apparent that a relatively broad energy loss feature is present and that it shifts to higher energy with increasing parallel wavevector.  
The position of the loss feature for several scattering conditions is shown in Fig. \ref{Fig_Be(0001)}. 
The slope of the dispersion curve turned out to be 5.5 eV $\rm \AA$ (for comparison: 1\,eV $\rm \AA \cong 1.52\times 10^5$\,m/s), 
slightly lower than predicted by a simplified 
1D model and in very good agreement with the prediction of a more accurate {\em ab-initio} model. Measurements along 
$\Gamma K$ were performed by another group \cite{Jahn2012}. The slope (group velocity) came out to be slightly higher 
(6.4 $eV \rm \AA$) than for the $\overline{\Gamma M}$ direction. Moreover along the $\overline{\Gamma K}$ high symmetry direction the 
ASP has a larger width than along $\overline{\Gamma M}$ due to the fact that the ASP dispersion is closer to 
the electron-hole pair continuum  along $\overline{\Gamma K}$ \cite{Silkin2008}.

\begin{figure}[tb]
\centering
\includegraphics[width=\linewidth]{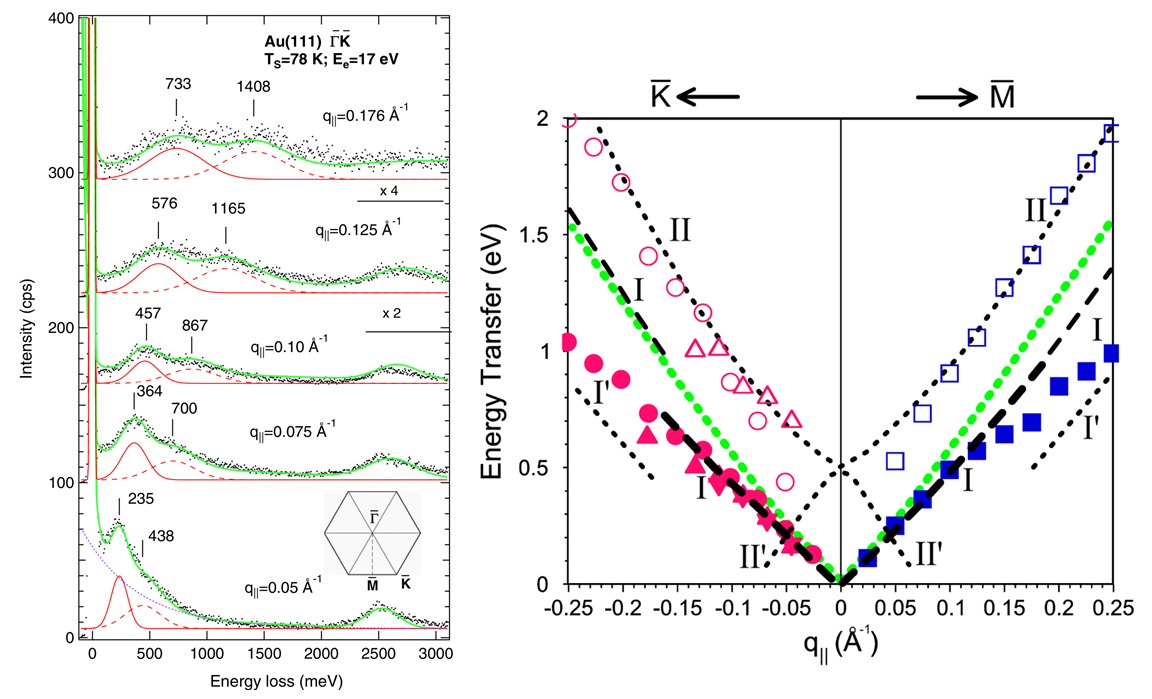}
\caption[]{Left: HREEL spectra recorded for different scattering conditions for Au(111). Right: dispersion 
curves for the lower and higher energy branches as measured along the two high symmetry directions are compared with the theoretical prediction. 
Reprinted with permission from ref. \cite{Vattuone2013}.
\label{Fig_Au(111)_Vattuone}}
% figg 2,4 \cite{Vattuone2013}
\end{figure}

The discovery of the ASP on Be(0001) triggered further measurements on other surfaces 
such as Cu(111) \cite{Pohl2010} and Au(111) \cite{Park2010}, for which theory 
had already predicted its existence. For Au(111), actually two linearly dispersing losses were found 
\cite{Vattuone2013} with group velocities of 3.95 $eV \rm \AA$ and 8.1 $eV \rm \AA$, respectively. 
Only the latter loss was seen in ref.\,\cite{Park2010}. A selection of spectra for different scattering 
conditions and the dispersion curves are shown in
Fig. \ref{Fig_Au(111)_Vattuone}. According to 
time dependent DFT calculations \cite{Vattuone2013}, however, only the first with the lower slope corresponds to the ASP, whereas the second  
higher energy branch was assigned to an interband transition 
between a band edge of the bulk and the  Shockley surface 
state. A further contribution to this loss from a multipole ASP cannot be excluded.  
 
%In the first measurement for Au(111) a linearly dispersing loss with a 
%surprisingly high slope (8.1 $ev \rm \AA$) was found and assigned to the newly discovered ASP \cite{Park2010}. 
%The large discrepancy between the theoretical prediction for the slope and this observation stimulated further 
%investigations of the same system. An experiment using an ELS-LEED setup confirmed the existence of the energy 
%loss feature observed by Park et al., but showed a further linearly dispersing band characterized by a 
%significantly lower slope (3.95 $eV \rm \AA$).\cite{Vattuone2013}.

Interestingly, as seen from Fig.\,\ref{Fig_Au(111)_Vattuone}, the slope of the low energy branch corresponds 
to a value of $\alpha <$1 in eq.~\ref{disp_asp}, 
i.e. to a loss running slightly below the edge of the electron - hole pair continuum. As demonstrated theoretically 
\cite{Vattuone2013}, this is a particular screening effect of the surface state excitation by the bulk electrons. 
For Au(111) the electrons in the Shockley surface 
state close to the Fermi level have velocities higher than the corresponding majority of bulk electrons, 
thus leading to a value of $\alpha$ slightly lower than 1. 
The complication with two losses also appeared for the Cu(111) case, as a recent investigation with very high energy resolution 
showed, i.e. the loss feature assigned to the ASP by Pohl et al. \cite{Pohl2010} consists actually of two losses. 
The component with the lowest slope among them was then assigned to the ASP for Cu(111), suggesting 
a scenario similar to Au(111) \cite{Pischel2013}.
%Spectra and dispersion found in this latest study are shown in Fig. \ref{Fig_Cu(111)}.
%\begin{figure}[htb]
%\centering
%\includegraphics[width=1\linewidth,angle=0]{Fig_Cu(111)_Pucci.pdf}
%\caption[]{Left: HREEL spectra recorded for different scattering conditions for Cu(111). 
%Off-specular HREEL spectra of Cu(111) exhibiting up to three broad loss features. The dark spots represent the measured data
%points, the (upper solid) blue lines are least-squares fits of two (three) Gaussians to the data. The different 
%contributions to the energy loss are shown as solid (excitation I), dashed (excitation II), and dotted (excitation III) 
%red lines, respectively, and shifted against the spectra for clarity. The spectra are not normalized to
%the intensity of the elastic peak. Note the different ordinate scales. The
%angle of incidence is $\theta$ = 60$^{\circ}$ for all spectra shown here. Incident
%energy $E_0$ and off-specular angle $\psi$ are indicated. Right: dispersion curves for the different contribution 
%to the loss peaks. The ASP is indicated by the green line. The dotted line shows the best fit to the dispersion 
%curve obtained in a former investigation \cite{Pohl2010} by some of us.
%Reprinted with permission from ref. \cite{Pischel2013} Copyright from ACS.}
%\label{Fig_Cu(111)} 
% figg 1,2 \cite{Pischel2013}
%\end{figure}
Thanks to the higher energy resolution, the width of the ASP is now lower than estimated before: using the time-energy principle 
of indeterminacy, it is thus possible to estimate the lifetime and the propagation length of the ASP, a parameter 
relevant to assess the feasibility of ASP-based devices. 

\subsubsection{ASP for nanostructured and regularly stepped surfaces}
After having shown that the ASP exists for most of the surfaces for which its existence had been predicted, a 
quite natural question arises: How robust is this excitation? Is it affected by surface nanostructuring?
In order to address such issues, experiments for both ion bombarded Cu(111) and for stepped Au surfaces were performed.
In the former case domes and pits of different size form upon ion bombardment, the morphology of which is 
determined by the interplay between thermal diffusion and removal of atoms by the incoming ions. 
In the latter case an ordered array of terraces separated by monoatomic steps is present.
The ASP turns out to be a relatively robust excitation:  it is not destroyed by surface nanostructuring 
both at RT and at 393 K, at least for moderate ion doses \cite{Vattuone2012}. 
The linear best fit to the dispersion data yields a slope of $(3.7 \pm0.5 eV\rm \AA)$. This value has 
to be compared with that found for the pristine surface ($(4.33 \pm 0.33) eV\rm \AA$, \cite{Pohl2010}).
As demonstrated by Theilmann et al. \cite{Theilmann1999} the Shockley Surface state (SSS) shifts from -0.4 eV to about 
-0.3 eV when the roughness of the Cu(111) surface is increased. This causes a decrease of the Fermi velocity 
and thus a decrease of the slope of the plasmon dispersion, in qualitative agreement with experiment.
For large enough ion doses and while sputtering at 393 K a non-dispersing loss was observed and tentatively 
assigned to a non-vertical inter-band
transition from the bottom of the SSS to the Fermi level, the required momentum being provided by surface 
disorder \cite{Vattuone2012}.
\begin{figure}[htb]
\centering
\includegraphics[width=0.5\linewidth]{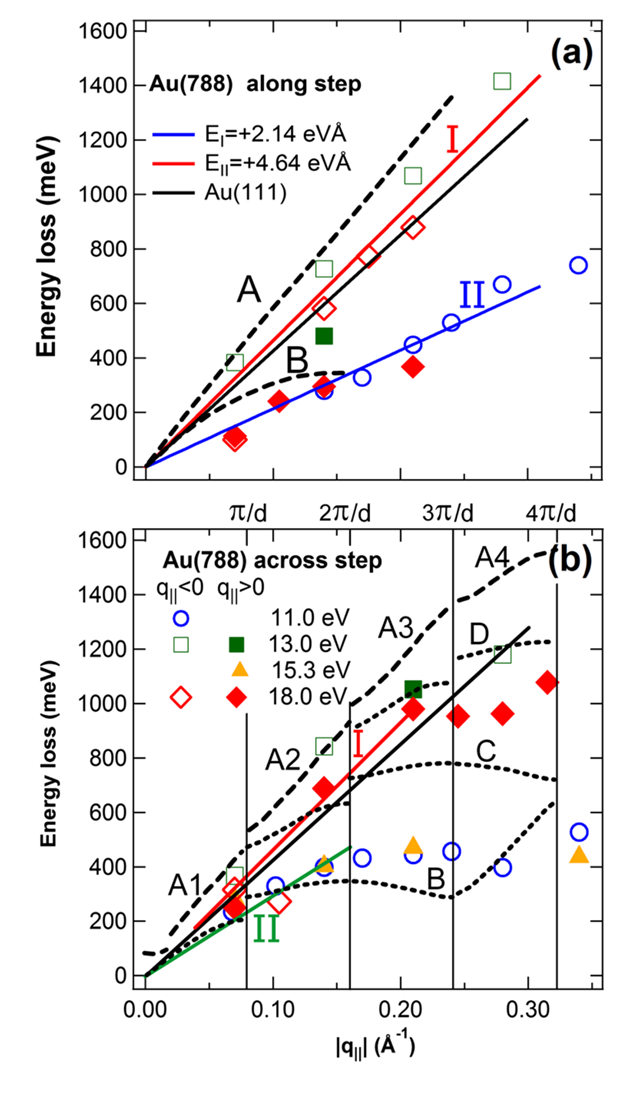}
\caption[]{Dispersion curves along and across steps: the symbols (see legend) indicate the experimental data points, the dotted 
lines the results of the theoretical calculations. The best fits to the experimental data is shown by light coloured lines. 
Reprinted with permission from ref. \cite{Smerieri2014} Copyright from APS.}
\label{Fig_Au(788)_Smerieri} 
% fig 4 \cite{Smerieri2014}
\end{figure}

With decreasing wavelength, also plasmons become more and more susceptible to scattering at atomic defects. 
This effect has clearly been seen for plasmons scattered at substrate steps in graphene on SiC(0001) \cite{Langer2010}. 
On a metallic surface, scattering at steps is expected to be smaller due to the less directed bonds, but as was demonstrated 
for the regularly stepped Au(788) surface, it leads to subband formation and partical localization of plasmons. 
The (788) surface consists of (111) terraces that are $\approx 40 \rm \AA$ wide, separated by monatomic steps.
The choice of this particular terrace size is motivated by the fact that the distance between steps exceeds 
the critical value (around $20 \rm \AA$ for Au) beyond which the short-range interaction between steps becomes negligible, 
so that opening of an energy gap in the surface band structure can occur \cite{Ortega2002}.
The choice of Au surfaces is motivated by their chemical inertness: the existence of the ASP is indeed closely 
related to the existence of a partly unoccupied Shockley surface state, which can in turn be destroyed by surface oxidation.
The dispersion of the ASP for Au(788) was measured using an ELS-LEED setup \cite{Smerieri2014}. 

The data are summarised in Fig. \ref{Fig_Au(788)_Smerieri}.
% It is immediately apparent that different 
%losses are measured at 11 and 13 eV \cite{Smerieri2014}. 
%Indeed, 
Two losses with sound-like dispersion are present, denoted by I
and II. The branch with the steeper slope originates from the ASP associated to the SSS of the (111) terraces, 
and has indeed a very similar slope as the ASP on flat Au(111). The branch with lower slope has no analogue on Au(111).
Although fully {\it ab initio} calculations are not feasible because of the too large unit cell, 
it was possible with the help of theoretical calculations  to assign all the features observed experimentally \cite{Smerieri2014}. 

In particular we underline that:
\begin{itemize}\item[(a) ] Along the steps two ASP modes exist. They have different group velocities
determined by the Fermi velocities of two partly
occupied SSS sub-bands.
\item[(b) ] The anisotropy introduced by the steps on Au(788) does not automatically lead to
plasmon localization normal to the terraces. 
\item[(c) ] The QW1 and QW2 sub-bands are still able to generate propagating plasmonic modes across the steps at wavelengths longer than the terrace width. The
slightly different slopes of the modes parallel and normal to the steps reflect the anisotropy of the system.
\item[(d) ] Across the steps, plasmon localization becomes visible when $q_\parallel$ exceeds the reciprocal lattice vector of the periodic step array.
\end{itemize}

The interest for the effect of the presence of a regular array of steps on the ASP existence and 
dispersion is not purely academic:
it has been suggested indeed that vicinal surfaces might provide naturally a diffraction grating of 
the nanometric size needed to couple the ASP with light \cite{Pitarke2007}. 
Due to the large difference between the speed of light and the Fermi velocity in the Shockley 
Surface State (almost 3 orders of magnitude!), an (infrared) photon with an energy of 
$\approx$ 0.5 eV could excite on Be(0001) an ASP with a wavelength of 50 to 80 $\rm \AA$, provided that 
a diffraction grating  allows to bridge the momentum gap.
A hypothetical device converting effectively photons into ASP should therefore have a characteristic spacing of that order, i.e. about two/three 
orders of magnitudes smaller than a traditional optoelectronic system operating at the same frequency. 

Although some theoretical estimates for nanoparticles predict a low efficiency for the conversion of photons into ASPs 
\cite{Hrton2012}, no experiments have been reported yet. 
In Table \ref{Table_slopes} experimental and theoretical values obtained for the slope of the ASP dispersion are summarised.

\begin{table}[tb]
\begin{tabular}{|l|c|c|c|c|c|c|r|} \hline
System & Fermi velocity ($\frac{e^2}{\hbar}$) & Experimental slope  & $\alpha_{exp}$ & $\alpha_{teo}$  \\ 
 & & ($eV\cdot \rm \AA$) & & (non ab initio: \cite{Silkin2005} ) \\ \hline
Be(0001) & 0.41 & 5.5 \cite{Diaconescu2007} & 1.08 & \\ \hline
Cu(111) & 0.28 & 4.33 \cite{Pohl2010} & 1.13 & 1.053 \\ \hline
Cu(111) & 0.28 & 3.11 \cite{Pischel2013} & 0.81 & \\ \hline
Au(111) & 0.35 & 3.95 & 0.83 & 1.032 \\ \hline
Ag(111) & 0.11 & - & - & - \\ \hline
\hline
Ag(111) & & & - &$\approx$ 7 \cite{Yan2012}\\ \hline
\end{tabular}
\caption{Fermi velocity, experimentally determined slope of the dispersion, experimentally and theoretical $\alpha$ for different systems.
The last row shows the theoretically predicted slope of the dispersion as estimated by the authors from Fig. 2c of Ref. 
\cite{Yan2012}.}
\label{Table_slopes} 
\end{table}

Summarizing the ASP observations, is apparent that the interaction between two electron gases provides a mechanism of 
mutual screening that leads to linearization of the dispersion, particularly also for small $q_\parallel$.
Therefore, it is not described by eq.\,\ref{eq1}, which at small $q_\parallel$ predicts a $\sqrt{q_\parallel}$-dependence. 

The fact that $\alpha$ can be $< 1$, as observed for Au(111), so that the ASP runs below the electron hole pair continuum,
does not directly lead to overdamping of the ASP, as often assumed. Possible reasons are small overlap between SSS and 
bulk states, but also phase mismatch of wave functions, expressed by the different Fermi velocities of bulk and surface state. 

The reliable calculation of these phenomena is still a challenge for simulations, as seen by two conflicting predictions for the Ag(111) surface: 
according to Ref. \cite{Yan2012} this system should 
exhibit a slope even higher than for Au(111), while the non ab initio theory predicts a slope even lower than on Cu(111), 
in agreement with a third, independent calculation \cite{Ahn2016}. 
No experiments have measured the ASP dispersion for Ag(111) yet, so that this controversy remains open.
On the other hand, the experiments carried out so far for ASP by HREELS measurements demonstrate that ambituities due to the 
appearance of several losses can only be resolved by reliable quantitative theoretical descriptions.

\section{Quasi-one dimensional plasmons}

As already demonstrated above for the case of Au(788), the introduction of extended defects such as 
steps on highly symmetric surfaces reduces the symmetry of surface states so significantly that subbands 
are formed, as also visible by two separate subband-plasmons for the two occupied subbands in the ground state.
These plasmons, however, propagate across the steps for long wavelengths and get only localized to single terraces for
wavelengths of the order of the terrace width. Therefore, they cannot be considered to be quasi-one dimensional. 

On the other hand, the experimental observation of a purely 1D dispersion of a plasmonic signal is clearly a necessary 
condition for quasi-1D behavior, but is it sufficient? In the ideal case, a Tomonaga-Luttinger liquid may appear, in which, due to 
spin and charge separation of the lowest excitations, the plasmon should appear as a holon with a linear dispersion \cite{Tomonaga1950, Luttinger1963,Voit1995}. 
While there is no final answer to this question yet, we will discuss 
in the following systems in which the requirement of 1D dispersion is cleary fulfilled: the formation of metallic chains of single atomic height
on regularly stepped Si surfaces. Regular arrays of these chains  are formed by self-organization, and the widths of the wires 
vary from single atomic chains to metallic strips that fill 
a whole (111) terrace, depending on the metal, but also on the step orientation. The main question is of course, 
how strongly the potential 1D properties of these wires are modified by their environment and the coupling to neighboring 
wires. As it turns out, there is again  a wide variation in the strength of interaction,
which still has to be understood.

A possible approach to 1D plasmonic properties start from a 2D quasi-free electron gas that is confined to a wire of finite width by an appropriate 
potential. Within this model, following ref.\,\cite{Moudgil2010}, the plasmon dispersion for a 
single isolated wire can be expressed as a function of the upper and lower boundary of the 
electron-hole continuum of excitations, $\omega_{\pm} = q_\parallel^2/2 \pm q_\parallel k_F/m^\star$ 
as 
\begin{equation} 
\omega_p(q_\parallel) = \sqrt{\frac{\omega_-^2 - \omega_+^2 e^{A(q_\parallel)}}{1-e^{A(q_\parallel)}}}
\label{eq1D}
\end{equation}
with $A(q) = 2\pi q/ (m^\star g_s V(q)[1-G(q)])$. V(q) is the Fourier transform of the confining potential, $G(q)$ the local field correction factor
due to electronic correlations, and $g_s$ the spin degeneracy (1 or 2). Already this model yields to 
lowest order a dispersion linearin $q_\parallel$, contrary to a 2D electron gas. As one sees from 
eq.\,\ref{eq1D}, the dispersion depends not only on electron density (via $k_F$) and effective 
masses, 
but also explicitly on the form of the confining potential, and electronic correlations. 
For the coupling between wires, no such analytic expression can be given, but an
approximate description of coupling, valid in the limit of small $q_\parallel$ \cite{Sarma1985,Li1990,Sarma1996} exist and will be 
discussed below for Au wires on Si surfaces. 

\subsection{Arrays of quasi-1D wires with small coupling}

From the few investigations of low-dimensional plasmons carried out on quasi-1D systems so far, 
Ag/Si(557) \cite{Krieg2013,Krieg2014,Krieg2015} and DySi$_2$/Si(100)vic. \cite{Ruge2010} are 
those systems in which the properties of individual wires  seem to be dominant, i.e. the coupling 
between wires is weak. Since the plasmonic properties of DySi$_2$/Si(100)vic. resemble those in 
the Ag/Si(557) system in many aspects, we concentrate here on the latter system. 

As mentioned above, the Ag/Si(111) system is a prototype system for the investigation of 2D 
plasmons, since carrier concentration in the Ag-modified Si surface state can be tuned by Ag 
surplus concentration. Similar properties have been found in the Ag/Si(557) system by 
investigations of the concentration dependent plasmonic properties. Although the doping mechanism 
on the flat and stepped surfaces need not be exactly the same, these studies allow to 
gain some surprising insights, not only into the doping mechanism, but also into general plasmonic 
properties in 1D \cite{Krieg2013,Krieg2014,Krieg2015}. A possible reason for the decoupling of wires on 
different (111)-oriented mini-terraces in this system may be the separation of these terraces by (112)-oriented
minifacets, which contain 3 atomic double-steps. 
\begin{figure}[tb]
\centering
\includegraphics[width=0.6\textwidth]{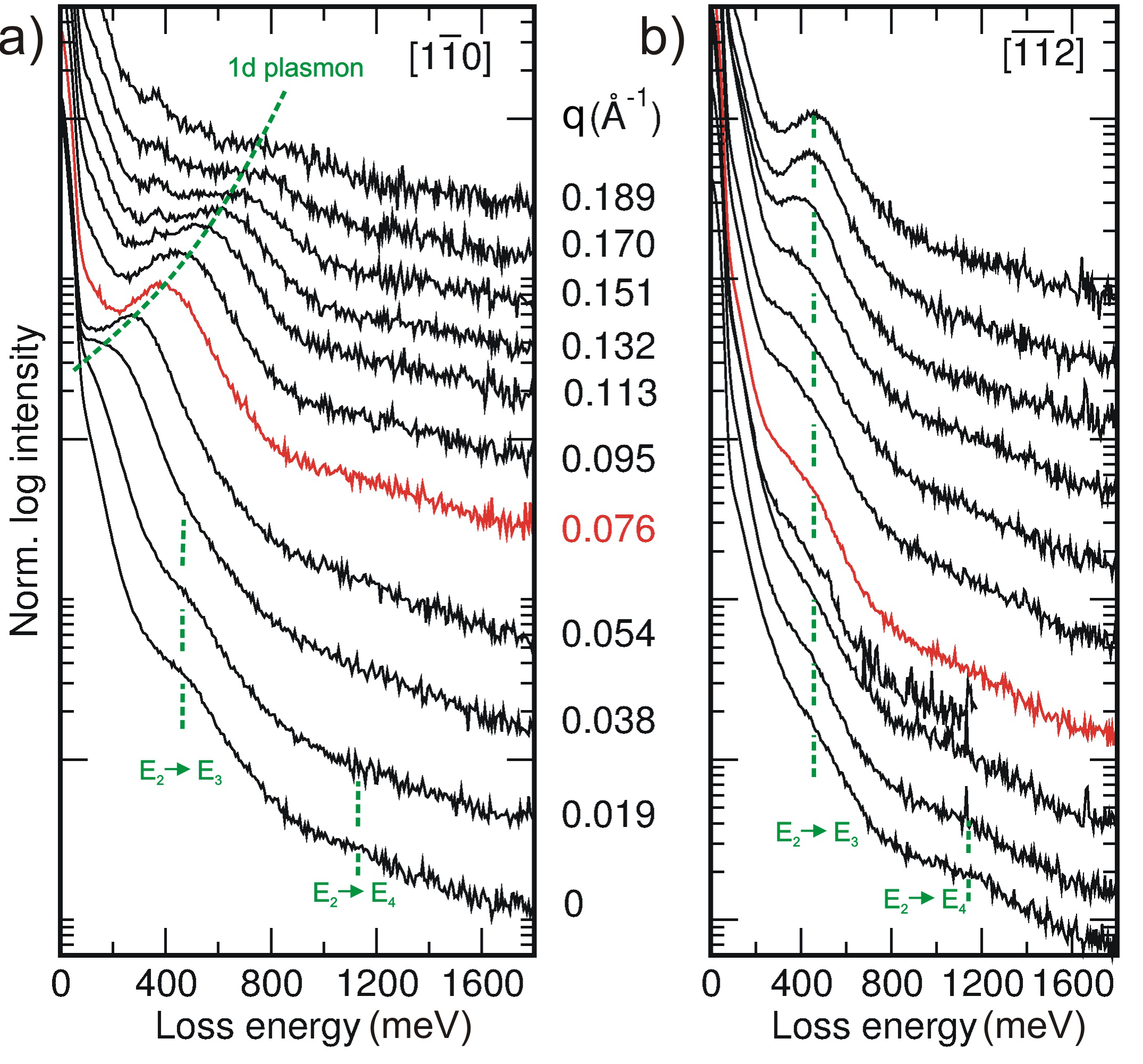}
\caption{EEL-spectra of 1 ML Ag/Si(557) along the $1\overline{1}0]$ (a) and $[\overline{1}\overline{1}2]$ (b) 
directions taken at 300 K. Apart from the dispersing 1D plasmon loss in a), two 
non-dispersing quantum well transitions E2 $\rightarrow$ E3 and E2 $\rightarrow$ E4 at 470 meV and 
1160 meV  are seen. Spectra are shifted for better visibility.
\label{Ag-steps-spectra}}
\end{figure}

\begin{figure}
\includegraphics[width=0.45\textwidth]{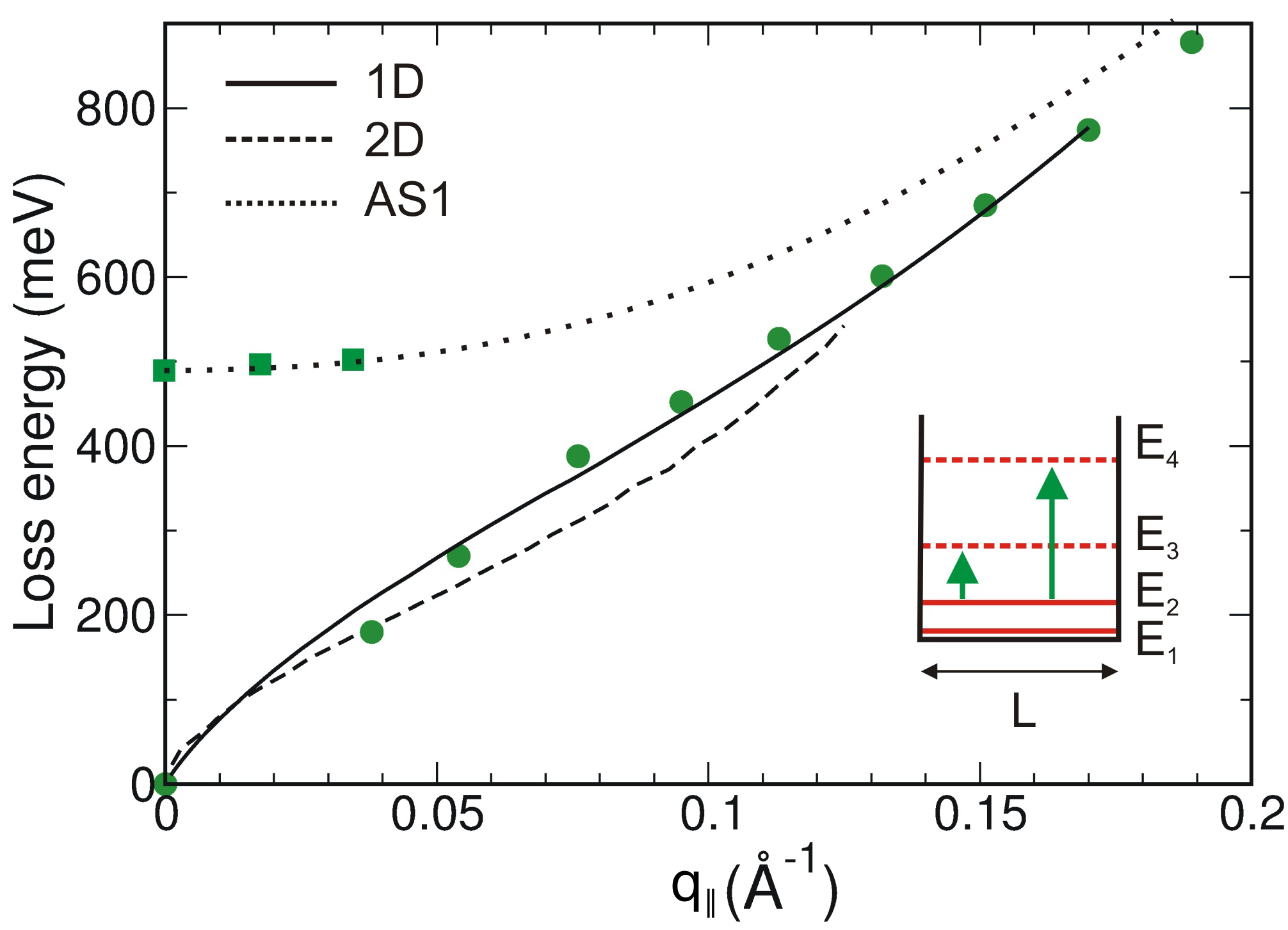} \hfill 
\includegraphics[width=0.33\textwidth]{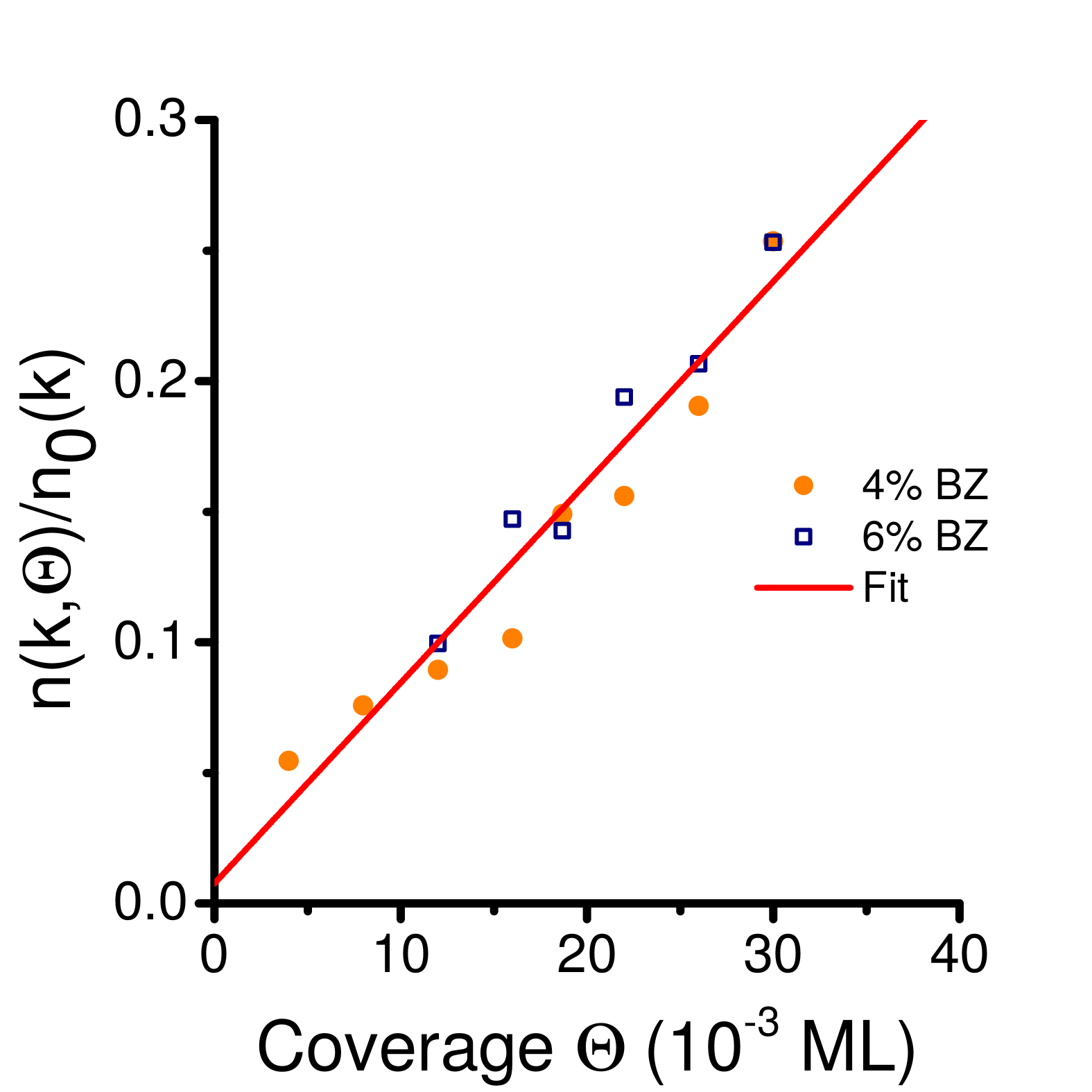}\hspace*{5ex}
 \caption{Left: Dispersion along the $1\overline{1}0]$ direction. 
For comparison the 2D plasmon dispersion in Ag/Si(111) (ref.\,\cite{Nagao2001}) is shown. The 
dotted line is a guide to the eye for AS1-mode. The AS2-branch is not shown.  
Right: Electron concentration, normalized to saturation, deduced from the position of the 
plasmonic losses as a function of Ag surplus concentration exceeding 1ML. \label{Ag-steps-disp}}
\end{figure}

The  atomic Ag wires formed in this system at coverages below 0.3 ML turned out to be  
semi-conducting. Low energy plasmons do not exist below 0.3 ML, therefore. The formation of 
metallic wires is  coupled to the appearance of $\mathrm (\sqrt 3 \times \sqrt 3 
R30^{\circ}$-order on the (111)-oriented mini-terraces, which have a local coverage of 1~ML.

As seen in Fig.\,\ref{Ag-steps-spectra}, a clear 1D dispersion  was measured by EELS-LEED, which 
means that the triple steps between the (111)-terraces act as insulating separators between the 
conducting strips.  Indeed, the measured dispersion for the lowest plasmon mode, shown in 
the right part of Fig.\,\ref{Ag-steps-disp}, agrees quantitatively with calculations 
\cite{Ruge2010}, which assume a single strip of a 2D electron gas that is confined by parabolic 
barriers to a width of 3.6 nm. Not only the 1D dispersion is reproduced, but also the appearance of 
intersubband plasmons along the wires, which are expected to appear as coupled 
(and simultaneous) excitations of quantum well states and plasmons in these wires of finite width 
\cite{Inaoka2005}. In the limit of long wavelengths they merge into the non-dispersing losses 
measured in the direction perpendicular to the wires (see Fig.~\ref{Ag-steps-spectra}b). They 
correspond to quantum well states formed normal to the wires. Their positions can be quantitatively 
fitted assuming a simple particle-in-a-box model with the measured wire width of 3.6 nm. 

\begin{figure}[tb]
\centering
\includegraphics[width=0.6\textwidth]{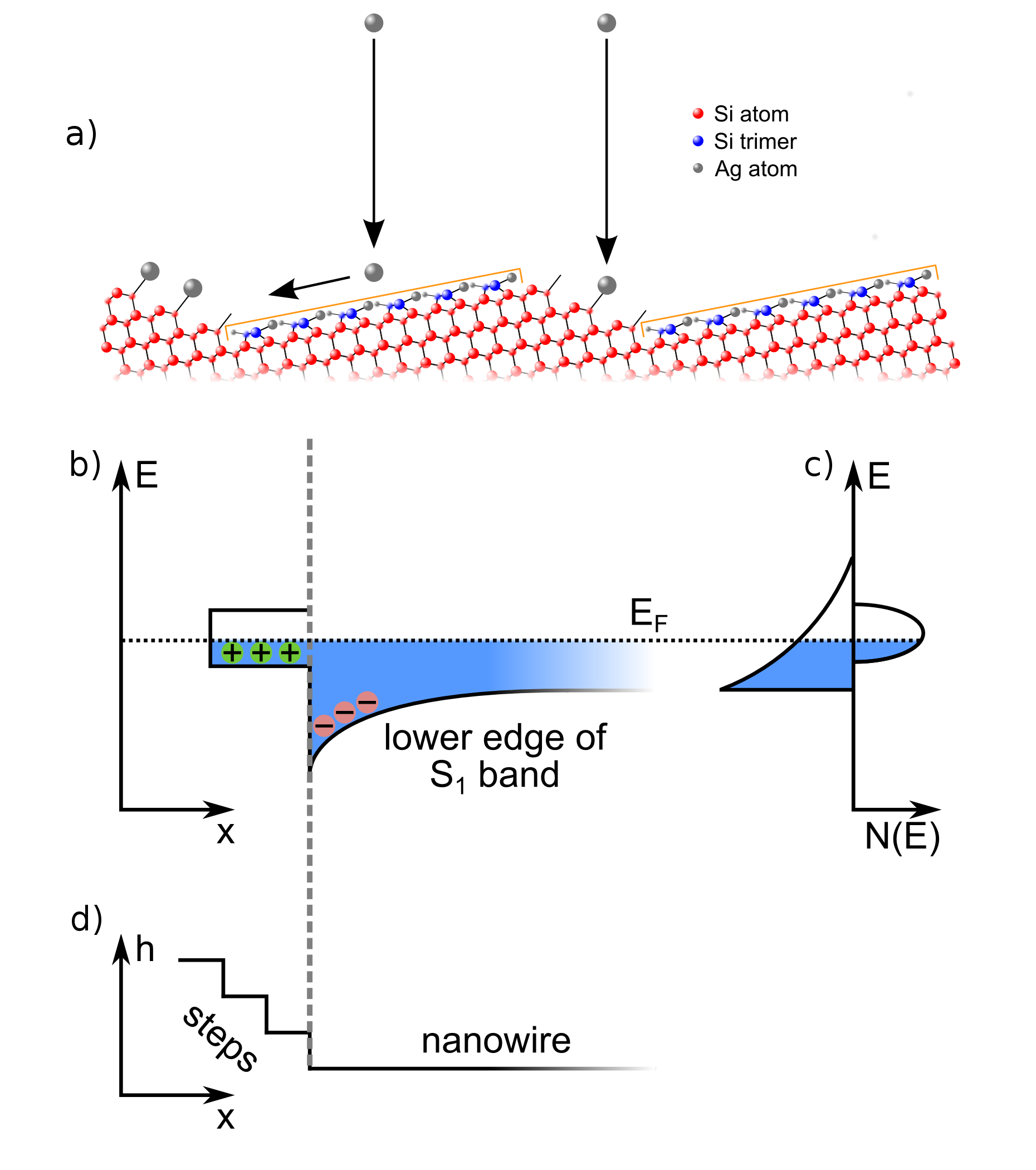}
\caption{a) Sketch of doping mechanism by adsorption of surplus Ag atoms (grey balls) on both the 
uncovered steps and on the Ag monolayer on flat terraces, where they diffuse and are trapped at 
step edges. b) Lateral band bending resulting from different adsorption energies (and 
corresponding level filling, see c)) at terraces and step edges, schematically shown in d).  
\label{Ag-doping}}
\end{figure}
The metallic properties of the wires are not intrinsic, but are induced by
extrinsic self-doping caused by Ag atoms adsorbed at the adjacent step edges, i.e. the bare 
Ag-wires with $\mathrm{\sqrt{3} \times \sqrt{3}}$-order are semi-metallic.  The doping atoms  can  
be desorbed thermally without destroying the majority of Ag wires on the (111)-terraces,
but can be made metallic again by the tiniest amounts of post-adsorbed Ag atoms. 
Thus this system allowed to determine precisely the dependence of plasmon excitation energy 
on doping concentration in the range between 0.003 and 0.03 of a monolayer. Assuming that all 
parameters except the electron density remain constant, and that each surplus Ag atom contributes 
the same amount of charge to the system at low concentrations, we found that  $\mathrm E_{plasmon} 
\propto \sqrt{n_e}$, in agreement with theory (see also right part of Fig.\,\ref{Ag-steps-disp}). 
Remarkably, this dependence turns out to be universal, independent of dimension, and to be valid 
even for a  Luttinger liquid. To our knowledge this is the first direct experimental verification of the 
$\sqrt{n_e}$ dependence for plasmons in a quasi-1D system.

This lateral doping mechanism on the atomic scale is most likely caused by the different binding 
strength of Ag atoms on terraces and at step edges. It leads to preferential trapping of Ag surplus 
atoms at the step edges and results in lateral band bending, as sketched in 
Fig.~\ref{Ag-doping}. As a consequence, the effective wire width is reduced with increasing 
doping concentration, since the doping chains of atoms are partially charged. This charging
also limits, depending on the chemical differences between terrace and step, the maximum doping 
concentration. Impurity atoms of different sorts can change this situation drastically, but this 
hasn't been tested yet in any detail. These experiments also clearly
ruled out the long debated doping process by a lattice gas in the second Ag layer, which turned out 
to be inactive with respect to formation of a low-energy plasmon.

The final proof for this model came from the fact that, similar to the Pb/Si(557) system (see below), also 
Ag/Si(557) modifies surface  and step energies so that at high temperature (here 600$^\circ$C)
the surface becomes unstable in the presence of a monolayer of Ag and tends to form larger (111) 
terraces and step bunches. Extended heating to 600$^\circ$C of the Ag covered 
surface roughly doubled the average terraces size. This allowed a 
unique proof of the extrinsic doping mechanism,
since the line doping concentration remains essentially constant  at saturation, but the 
overall concentration is reduced with increasing terrace size, since the mount of electrons donated by 
the dopant atoms is distributed over the larger (111) terraces \cite{Krieg2015}.

\subsection{Between 1D and 2D: Pb/Si(557)}
\begin{figure}[tb]
\begin{center}
\includegraphics[width=0.6\textwidth]{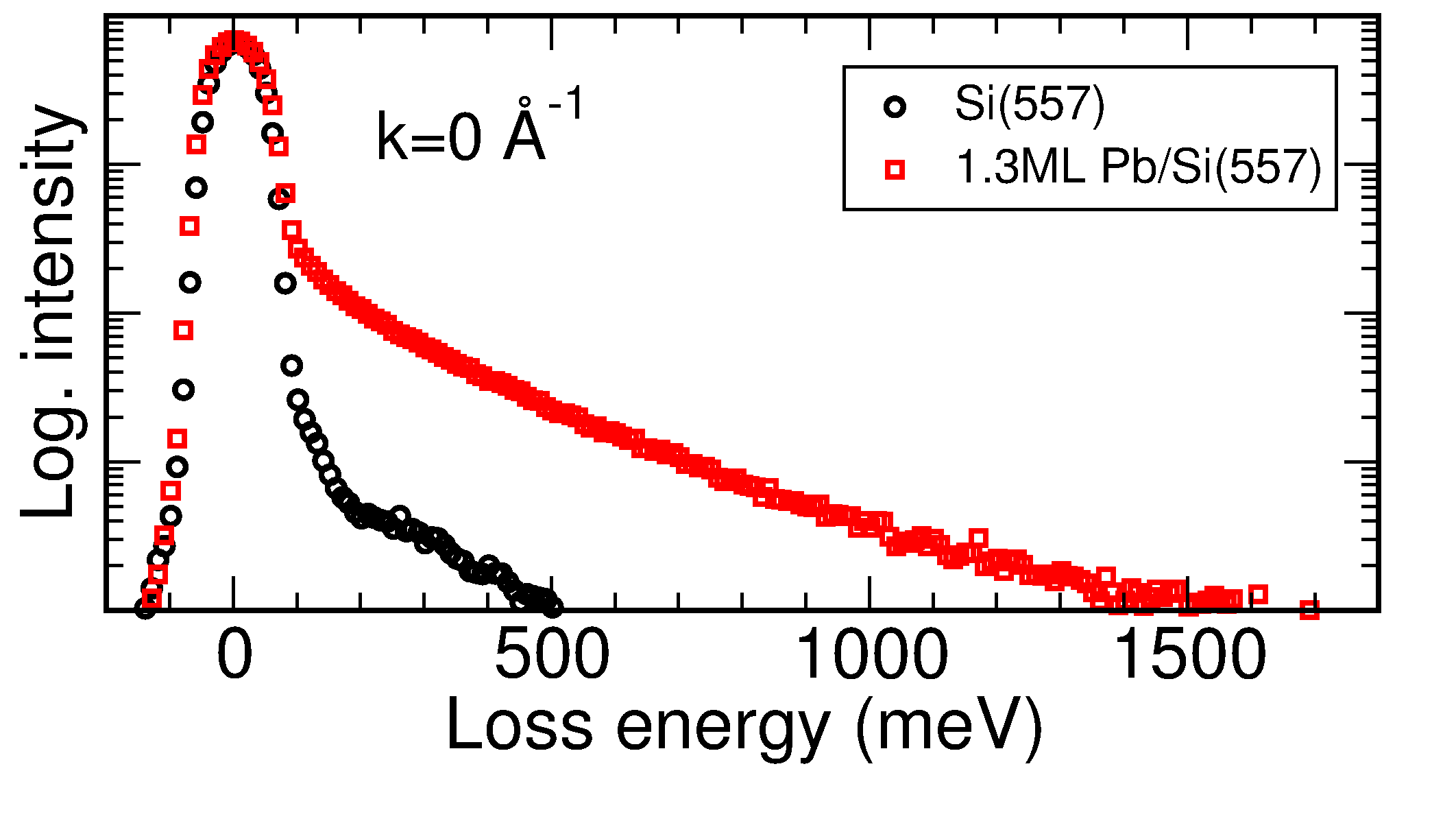}
\end{center}
\caption{Comparison of the electron loss spectra on clean Si(557)(circles) with that with 1.31 ML of Pb adsorbed at 
$q_\parallel = 0$. Due to the metallic property of the Pb layer, an exponentially 
decaying continuum of losses is seen (Drude tail). Primary energy $E_0 = 20$ eV.  
\label{eels}}
\end{figure}

The situation is much less clear in the Pb/Si(557) system, although at first glance it looks quite 
similar to Ag/Si(557). 
For a critical coverage of 1.31 ML and at low temperature (below 78 K) the system clearly has 1D properties in 
electrical 
conductance due to Fermi nesting and 1D band filling induced by facet formation. Facettiing locally 
transforms the surface 
into a (local) (223) orientation  \cite{Tegenkamp2005, Tegenkamp2008} that causes the opening of a 
small 1D gap 
of 25 meV around the critical coverage. 
On the other hand, this high Pb concentration exhibits relatively strong electronic coupling between 
the terraces 
so that the general appearance of the band structure is 2D-like,  although with 1D properties close to the 
Fermi energy.  
This leads to the  following intriguing questions: are such 1D properties still 
visible in collective electronic excitations with excitation energies by far higher than the bandgap of 25 meV ?
There  may even be a transition between 1D and 2D behavior 
depending 
on the available decay processes into single particle excitations.   

Metallicity of the layer at 1.31 ML Pb coverage can easily be identified even qualitatively 
by comparing the electron energy loss spectra for the clean and Pb covered Si(557) surface. 
Whereas for Si(557) only phonons can be created at small excitation energies, 
the continuum of low energy electronic excitations in a metallic system results in a broad 
structureless background with exponentially decaying intensity as a function of loss energy. 
This is a clear signature of metallicity for the monolayer Pb/Si(557) system.  
\begin{figure}[tb]
\centering
\includegraphics[width=0.45\textwidth]{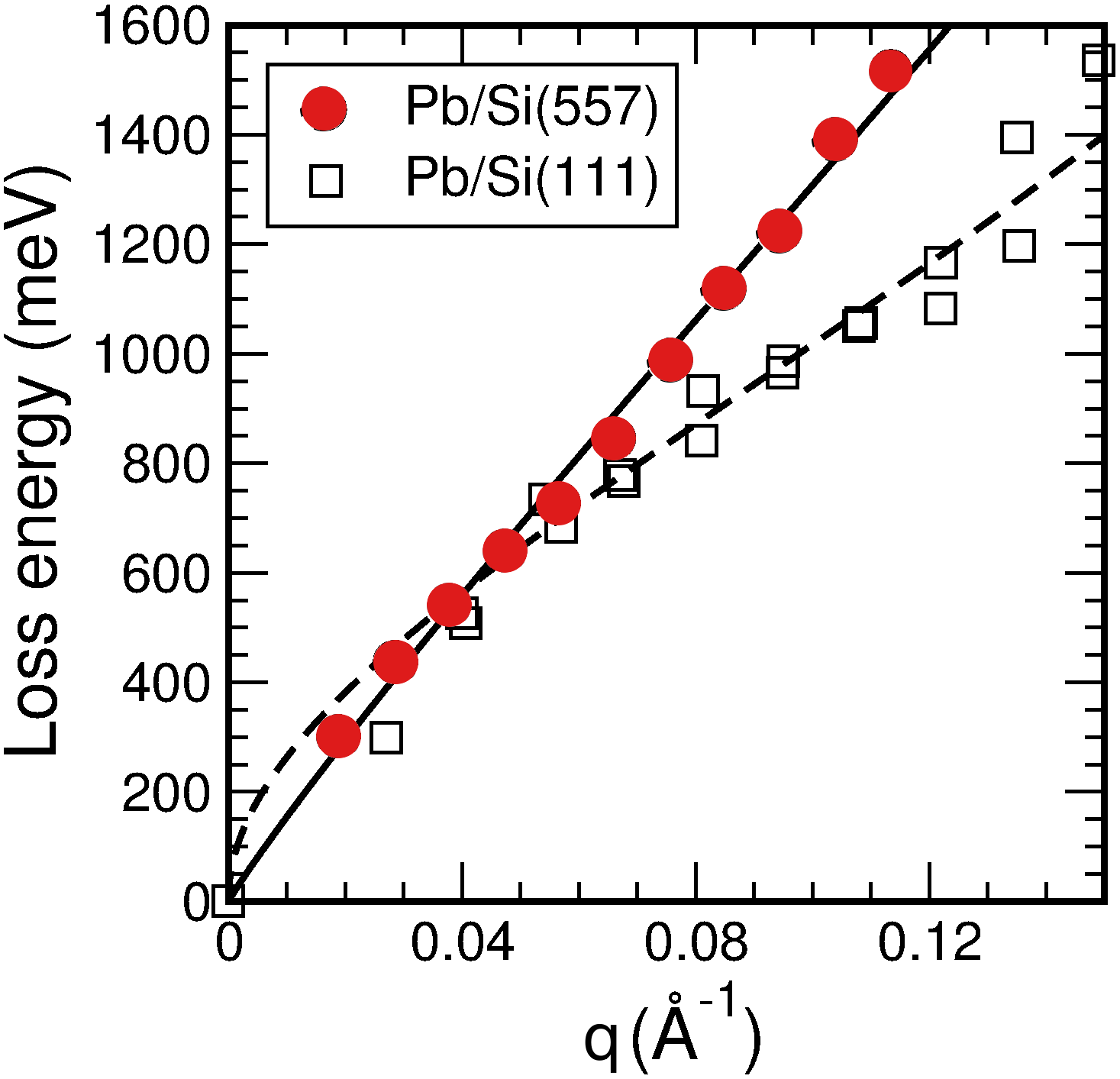}\hspace*{3ex}
\caption{Dispersion for 1.31 ML Pb on Si(557) parallel to the steps (red balls), compared with that 
obtained
for the same coverage of Pb on Si(111). The solid
curve is a quasi-1D fit with $m^\ast = 0.4m_e$ and N$_{1D} = 3.6 \times 10^7$~cm$^{-1}$ (see text), 
whereas 
the dashed curve is a 2D fit with the nearly free electron gas (NFEG) model with an effective mass 
of 
$0.4 m_e$ and an electron density of N$_{2D} = 2.4 \times 10^{14}$~cm$^{-2}$.
\label{disp-131}}
\end{figure}
 
For the critical Pb concentration of 1.31 ML the HREELS spectra 
as a function of $q$ both parallel and perpendicular to the terraces were measured  at room 
temperature and with $\ell$He cooling (Fig.~\ref{eels}). No clear temperature dependence was found. 
While 
in the direction along the wires a clear dispersing loss peak was observed, no such loss was 
detected in perpendicular direction. In other words, plasmonic 
excitations have clearly a 1D dispersion at this Pb coverage. 
As seen from Fig.~\ref{disp-131}, the measured data points can be naturally extrapolated to zero 
momentum and zero energy. The dispersion can quantitatively be described by a model of a 
confined 2D electron gas with local field corrections and correlations 
\cite{Inaoka2004,Inaoka2005} using a {\em single}, four atoms wide metallic strip that is confined 
in a 
parabolic potential with the effective mass, $m^\ast$, and the 1D electron density, $N_D$, given in 
the caption of Fig.\ref{disp-131}. Further details can be found in ref.\cite{Block2011}. 
As seen also by the comparison with an analogous measurement for the isotropic Pb/Si(111) layer 
(see Fig.~\ref{disp-131}), there is a clear qualitative difference between a 1D and a 2D system 
(note that both dispersions must go through zero). Contrary to the 
Ag/Si(557) system \cite{Krieg2013}, however, no indications for quantum well confinement were 
detected. 

The missing temperature dependence of the 1D plasmonic dispersion can be taken as 
an indication that the physical mechanism for decoupling the plasmonic excitation on the 
various terraces cannot be related to Fermi nesting and opening of the 1D band gap alone. 
The absence of quantum well states means that the simple picture of a confining potential must be 
questioned.
The key for this at first glance contradictory behavior may be hidden in the spin structure and the 
preferentially anti-ferromagnetic spin orientation on adjacent terraces, which leads to orthogonal 
electronic
states. While there are no measurements, part of this spin correlation may still exist even at room 
temperature and 
reduce coupling. In the direction normal to the steps, the high 
resistance of steps \cite{Roenspies2010}, which at this Pb concentration turned out not to be 
covered 
by Pb \cite{Czubanowski2008}, is already sufficient to prevent formation of plasmon waves in this 
direction. This 
situation, however, is modified by the gradual filling of step edges with Pb at higher Pb 
concentrations. 

\begin{figure}[tb]
\begin{center}
\includegraphics[width=0.46\textwidth]{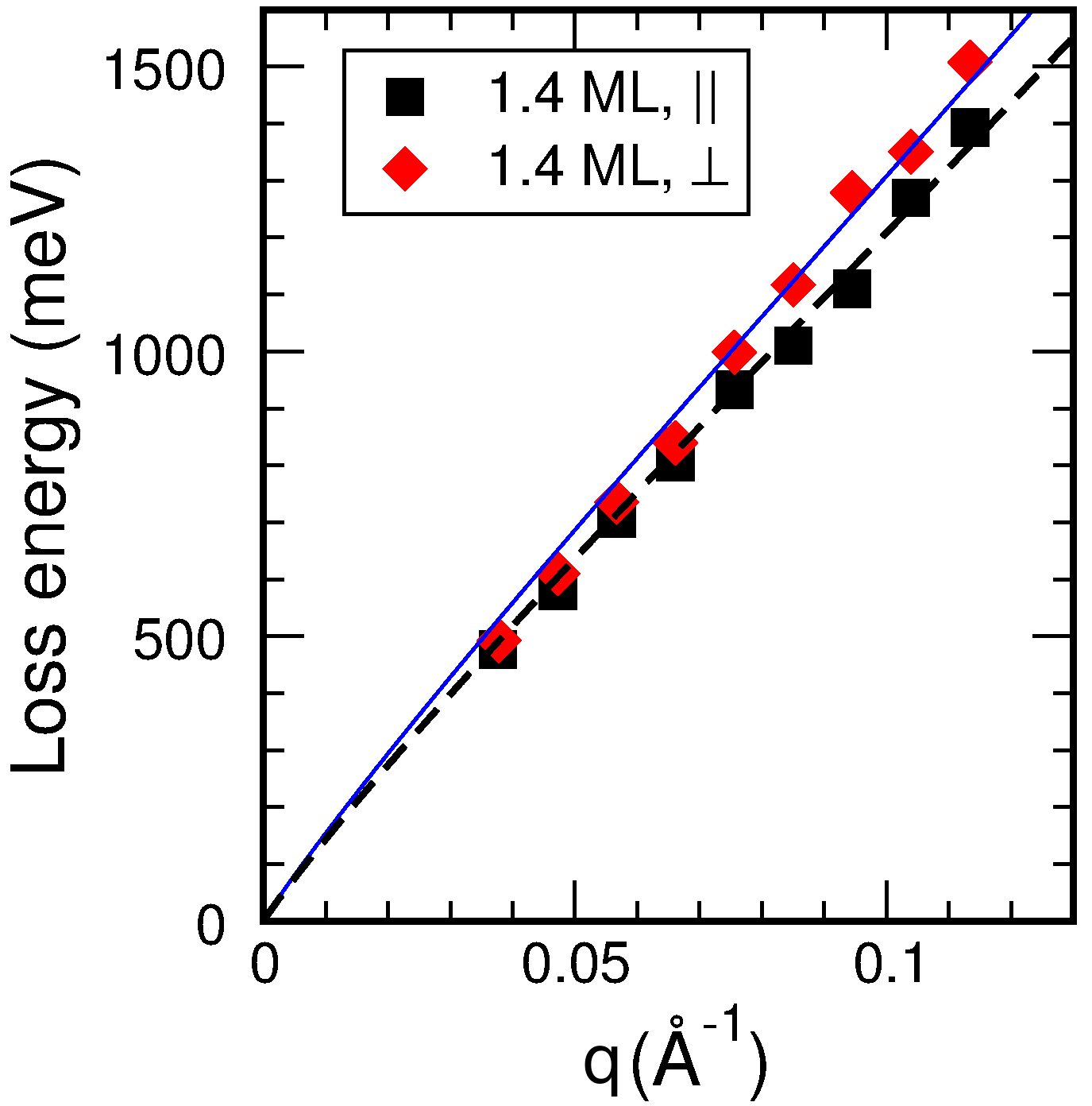}\hspace{2ex}
\includegraphics[width=0.41\textwidth]{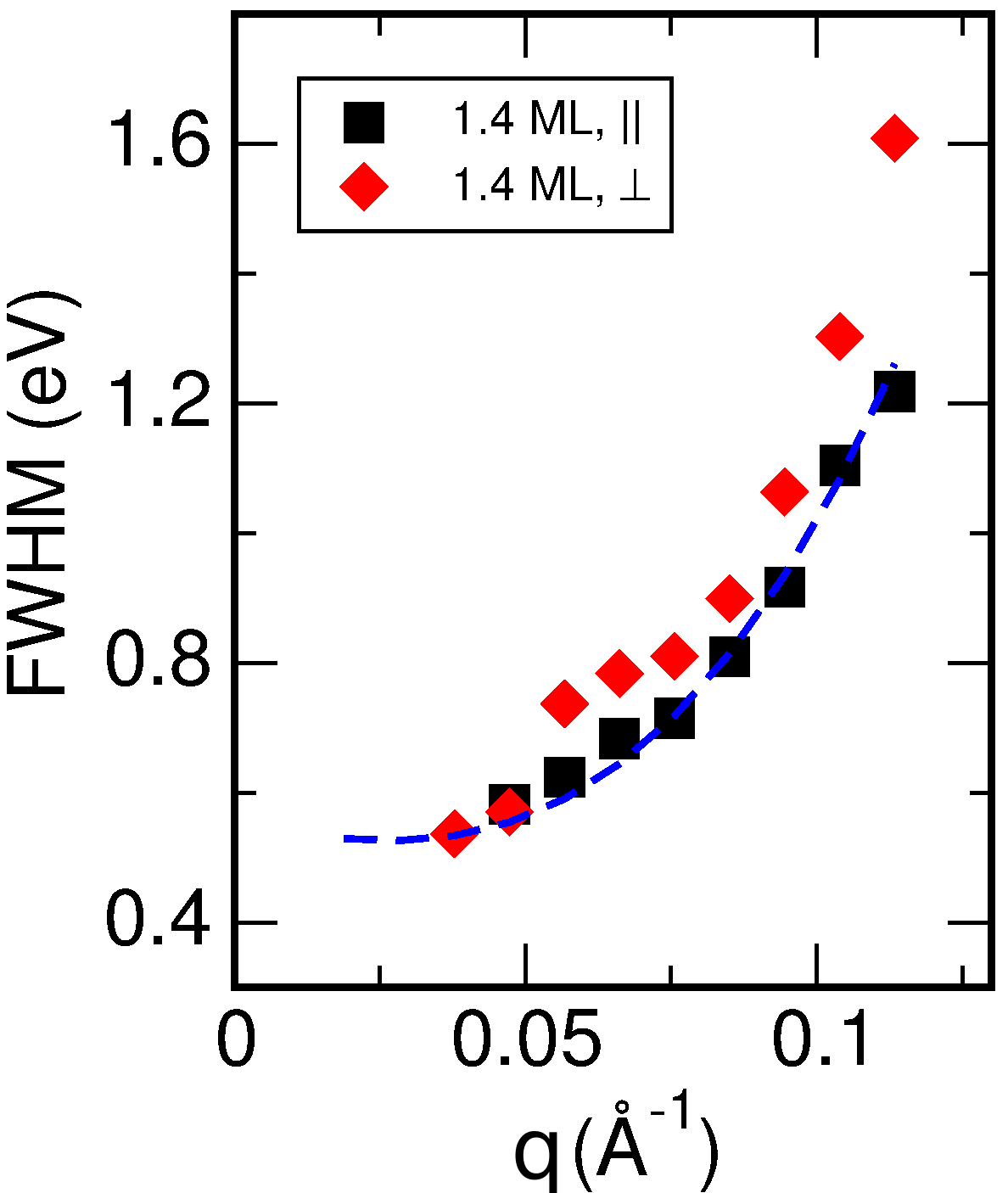}
\end{center}
\caption{left: Dispersion parallel and perpendicular to the Si(557) steps. Right: corresponding half 
widths (FWHM) 
of the measured losses. The dashed curves correspond to the results of 1.31 ML.
\label{disp-14}}
\end{figure}
Interestingly, the quasi-linear dispersion remains when the steps are decorated by Pb, as shown now 
by 
increasing Pb coverage to 1.4 ML. At this Pb concentration about half of the Si step edges are 
decorated
\cite{Czubanowski2009}. While these added chains increase coupling between different terraces, also 
the 
effective band gap is reduced \cite{Tegenkamp2010}. 
Both changes in coupling parameters between terraces 
obviously now lead to the behavior documented in Fig.~\ref{disp-14}. While there was no detectable 
signal 
for long wavelengths, losses and their dispersion were found both parallel and perpendicular to the 
steps 
for wavelengths shorter than 15 nm, and with a quite small anisotropy of about 10\%. 
This means that when changing the coupling between terraces, dispersing 
plasmonic excitations appear with wavelengths that exceed up to ten times the terrace width. 

It should be noted that in both directions still a quasi-linear dispersion is detected, similar to 
the Au(788) system 
discussed above. Anisotropy seems to be sufficient for linearization of the plasmon dispersion. This 
finding 
may be obvious from eq.\,\ref{eq1D} for the direction parallel to the wires, but not in 
perpendicular direction. 
Contrary to the Au(788) surface, we find here that propagating and dispersing plasmons normal to the 
wire direction are
found only at wavelengths considerably shorter than the undistorted periodicity of (223) facets. 
This lack of 
periodicity for distances larger than 20 nm of course also makes it impossible to form long 
wavelength plasmons here.
 
Thus there is a clear crossover behavior from 1D to 2D induced by a very small change of surface 
concentration of Pb. At this moment 
there is neither a clear understanding  of the underlying physical mechanism for 
the crossover nor  a quantitative theory for the plasmonic excitations in these anisotropic 
low-D systems. 

Interestingly, the metal-insulator transition of In/Si(111), historically the first well 
investigated quasi-1D system \cite{Yeom1999}, was seen not only in DC conductance, but 
could also be detected by the disappearance of the plasmonic loss \cite{Nagao2006,Nagao2007}, 
in contrast to the Pb/Si(557) system. 

These findings seem to be contradictory at first 
glance: the phase transitions in both systems are obviously of first order, though with 
opposite sign of temperature, the main difference lying in the structural changes involved. 
In Pb/Si(557) the Pb structure on each terrace essentially remains the same when going through the 
phase transition \cite{Pfnuer2016}. Just the average step distance is changed by only 
3.5\% \cite{Czubanowski2008}, which turned out to be enough to destroy the nesting condition at the 
critical coverage of 1.31 ML, and to form a sequence of charge density waves with varying 
periodicity and decreasing band gap normal to the steps as a function of increasing Pb 
concentration \cite{Czubanowski2009}. Also the electronic changes 
in the bandstructure of this system associated with the metal-insulator transition seem to be 
small.  The plasmon excitation seems to be quite insensitive to these small electronic changes. 
This scenario contrasts very much with the strong structural and electronic changes in the 
In/Si(111) system when going from the $(4\times 1)$ to the low-temperature $(8\times 2)$ phase 
\cite{Wippermann2010,Wall2012}, which involves melting and disappearance of the 
whole CDW system.

\subsection{Coupling in atomic wires: Si(hhk)-Au}

Au chains on regularly stepped Si(111) surfaces, 
tilted at various angles towards the \per direction, provide the narrowest possible arrays of 
chains consisting either of one or two atoms, depending on the type of step. Here only 
double steps of Si separate the (111)-oriented terraces. Although purely 1D dispersion along the 
chain direction is found, the lateral extension of the charge distribution, determined 
both by the structural motif and the terrace width, turns out to 
explicitly influence the slope of the measured plasmon dispersion curves. In other words, the
crossover into the second dimension is crucial for the quantitative interpretation of this 
quasi-1D phenomenon, but is not described completely by existing theories. 

Depending on coverage and vicinality, the widths of the Au-chains and their interwire 
spacing can be tuned, while their electronic band structures are still very similar. 
0.2\,ML of Au on Si(557), e.g., result in growth of single atom Au-chains and a row of Si-adatoms 
on each mini-terrace with an interwire spacing of 19.2\,\AA\ \cite{Crain2004,Nagao2006}. In 
contrast, Si(553) and Si(775)-Au 
host double Au chains in the center of the terrace \cite{Aulbach2016,Krawiec2010}. 
The interwire spacing is 14.8\,\AA\ for Si(553)  and 21.3\,\AA\ for Si(775) 
\cite{Crain2004,Barke2009,Erwin2010}. For double Au chains, 
nominal coverages of 0.48\,ML on Si(553) and of 0.32\,ML on Si(775) result.  
Common to all these structures is a graphitic Si-honeycomb chain located at the step edges 
\cite{Crain2004,Barke2009,Erwin2010}.
Each of these system is characterized by metallic bands that are well known from angular resolved 
photoemission (ARPES) measurements \cite{Crain2004}. 
They only disperse along the chain direction $q_\parallel$, and have their minima at the zone 
boundary. Thus, also the (equilibrium) electron density available for plasmonic excitations is well 
known.

Here we compare the collective excitations in Si(553)-Au and Si(775)-Au, which have the same 
structural motif of the double gold chain. Furthermore, the Si(553)-Au system allows to vary 
coupling between wires, because it forms two phases. In the high coverage phase 
(HCW) at a Au coverage of 0.48 ML all terraces are covered with a double strand of Au, whereas 
in the low coverage phase at a coverage of 0.21 ML only every second terrace hosts the double 
gold chain \cite{Song2015a,Lichtenstein2016}. We will also make reference to the 
Si(557)-Au system with only a single gold chain. 

Interestingly, the step structure also seems to have significant influence on the coupling between 
wires on different terraces. While the Au wires on Si(553) and Si(775) seem to be strongly coupled, 
as shown below, this coupling turned out to be weak for Au/Si(557) \cite{Moudgil2010}. 
The plasmonic coupling between the wires in the ordered arrays, which is another 
aspect of dimensional crossover, is quantitatively modeled by existing mesoscopic theories 
\cite{Li1990,Sarma1996}.

Contrary to the EELS spectra for Ag/Si(557) shown above, the spectra for all Au systems close to 
$q_\parallel = 0$ are structureless, apart from a small 
non dispersing feature that dies out quickly with increasing $q_\parallel$.
Metallicity is demonstrated by the exponentially decaying background intensity as a function of 
loss energy, known as Drude tail \cite{Krieg2015}, in agreement with 
findings from ARPES for the Si(553)-Au \cite{Crain2004}, STM \cite{Aulbach2013}, and theory 
\cite{Erwin2010}. However, this result is at variance with the ARPES data for the Si(775)-Au 
\cite{Crain2004} for reasons still to be explored. In the direction along the wires, clear loss 
features are observed, which shift to higher loss energies with increasing scattering angles, i.e. 
with  increasing $q_\parallel$. In the $k_\bot$ direction, however, and similar to Ag/Si(557), no 
dispersing mode is seen. 
\begin{figure}[tb]
\centering
\includegraphics[width=0.6\textwidth]{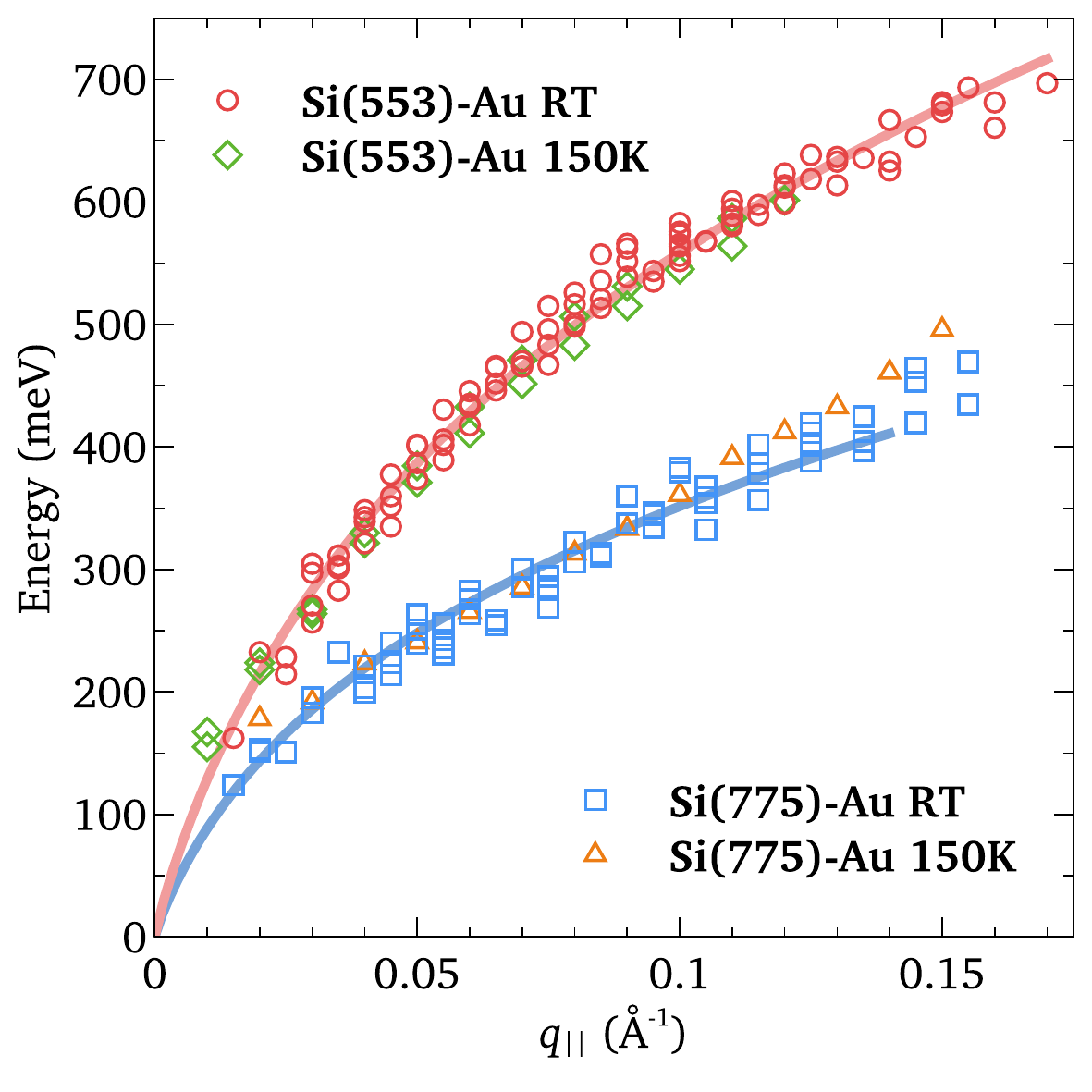}
\caption{Plasmon dispersion for Au quantum wires grown on Si(553) (HCW phase) and on Si(775). 
Lines are fits according to eq.\ref{eq2}. Dashed-dotted lines: first terms of 
eq.\,\ref{eq2} for both systems. Dotted line: same for Si(557)-Au, fitted data from 
ref.\,\cite{Nagao2006}.\label{disp-2}}      % 
\end{figure}

The  dispersion curves along the wires, derived from the loss maxima, are shown for Si(775)-Au and 
for the HCW phase of Si(553)-Au in Fig.\,\ref{disp-2}. From the ARPES data cited above, the ratio 
of electron densities and effective masses turns out to be virtually identical in the two systems 
for all surface bands. Therefore (see formula below), and in spite of two existing bands, only one 
plasmon loss is expected, in agreement with our findings.  Thus for a true 1D system, no 
dependence on terrace width, d,  is expected to first order. This is obviously not the case, since 
a 1/d dependence is found. 
Using existing theories for a nearly-free electron gas confined into quasi-1D wires 
\cite{Sarma1985,Li1990,Nagao2007,Inaoka2007}, a quantitative fit turned out to be possible with a 
modified model of coupled wires, sitting in a periodic array of square potentials at 
distance $d$ \cite{Li1990,Sarma1996}. At small $q_\parallel$, the dispersion is given by
\begin{multline}
E = \hbar \sqrt{\frac{4 n e^2}{(1+ \epsilon)\epsilon_0 m^* a^2}} q_\parallel a_0 \times 
\sqrt{\text{K}_0({\frac{q_\parallel a}{2 \sqrt{2}}) + 2 \sum_{l=1}^L \text{K}_0(q_\parallel 
ld)\cdot\cos{(k_\bot l d)}}} \label{eq2}
\end{multline}
where the first term of the product contains the electronic and structural properties of a single 
wire, the second the intrawire  (first term under square root) and the interwire interaction.
$n$ is the electron density per unit length, $e$ the elementary charge, $m^\star$ the effective 
mass. $\epsilon$ is the dielectric function of Si as partially embedding medium.
$\text{K}_0$ are modified Bessel functions of zeroth order and second kind, $k_\bot$ is the 
momentum normal to the wires.  
If $a$ (the effective wire width) is set equal to $a_0$ (a constant for normalization), 
eq.\,\ref{eq2} corresponds to the original formula given in 
refs.\,\cite{Li1990,Sarma1996}, which, however, turns out not to describe our findings.
The ratio $a_0/a$ accounts both for differences 
in structural motifs and effective wire widths of a single wire, and is the only free parameter in 
eq.\,\ref{eq2}.  In the array of square potentials the first term under the second square root 
accounts for the self-interaction of a single wire, whereas the second term describes the 
interaction between different wires at multiple distances of $d$.

A further test to the sensitivity of the dispersion to the structural motif was carried out 
by fitting the published data of ref.\,\cite{Nagao2006} for Si(557)-Au to eq.\,\ref{eq2}. The 
Si(775) and Si(557) surfaces have almost the same terrace widths (21.2 vs. 19.8\,\AA, 
respectively). 
For the best fit the first term of eq.\,\ref{eq2} has to be a factor of 1.6 {\it larger} for 
Si(557)-Au than for Si(775)-Au. 
Taking the differences in $n$ and the $d$-dependence from above for the two systems, the effective 
width $a$, as suggested  in eq.\,\ref{eq2}, 
is reduced by roughly a factor of 2 for Si(557)-Au compared to Si(775)-Au. 
These results demonstrate that not only the periodicity, given by the wire 
distances $d$, has a direct influence on the dispersion, 
but also the {\em internal 2D distribution} of the (excited) electron density within each wire. 
This result for Au/Si(557) is very much in agreement with the findings of ref.\,\cite{Moudgil2010}.
Subband excitations, as in the Si(557)-Ag system, can be ruled out here as possible source 
for the observed differences, since for these narrow structures and the given k$_F$ 
only the lowest subband of a quantum well is occupied. 
These results demonstrate that even in case of purely 1D plasmonic dispersion, there is a 
crossover to 2D, and both the width of a wire on the atomic scale and the internal electronic 
distribution within the very wire enter directly the slope of plasmonic dispersion. More details, 
e.g. about the choice of $a_0$, can be found in ref.\,\cite{Lichtenstein2016}.

As it turns out, the LCW phase of Si(553)-Au can be described by the same formalism of 
eq.\,\ref{eq2} \cite{Lichtenstein2016-2}. In fact, according to the analysis of Song et al. 
\cite{Song2015a}, the Au covered terraces are supposed to have the same terrace width as in the 
HCW phase, whereas the empty terrace acts as separator between the Au wires.

This same effective wire width, however, 
does not reproduce the measured dispersion for the LCW phase, if we only assume reduced 
coupling between wires due to their increased interwire distance from 
14.8\,\AA\ to 32.7\,\AA. 
By actually fitting the data with $a$ 
as a free parameter we obtain an effective wire width of 8.8\,\AA, compared to 
7.5\,\AA\ for the HCW phase.
This effective spreading of the plasmon excitation on the terrace is in fact quite 
plausible. In the LCW phase  
terraces not covered by gold are under tension \cite{Song2015a} with the consequence of atomically rough edges. This 
gives the plasmon mode the possibility to spill out more into the uncovered neighboring terrace by 
approximately $0.5a_\perp = 1.66$\,\AA.  Assuming the same proportionality factor between terrace 
width and effective wire width as for the 
HCW phase \cite{Lichtenstein2016}, a corrected value of 8.5\,\AA\ is obtained that is 
very close to the value obtained here. Therefore, this result fits into the general picture for 
the stepped Au covered Si surfaces that the electron density 
spills out as far as possible limited by the structure of the step edges, and 
expresses the crossover of two-dimensional properties on the one-dimensional plasmon dispersion.
The effect of spillout of electrons in the LCW phase compared to HCW may be also taken as 
indication of softening of the confining potential. A softer potential than the square would in 
fact have a qualitatively similar effect at constant effective wire width. 

Summing up this section, we demonstrated that the Si(hhk)-Au systems  offer the unique 
possibility to study  both the spreading of the chain-induced electron density on the 
mini-terraces and the influence of wire coupling in plasmonic excitations of quasi-one dimensional 
arrays of gold chains. As it turns out, there is an intriguing interplay between geometric 
terrace width and structural motiv that influences the effective wire width, as 
obvious from a comparison of, e.g., Si(775)-Au and Si(557)-Au. Coupling between wires, on 
the other hand varies both with terrace size and with the step structure. It is mainly the large 
average separation between terraces in the Si(557)-Ag system that makes them appear as essentially 
isolated wires. Weak coupling across Si(557) steps, which was found to be much weaker 
for Si(557)-Au than for Si(775)-Au, acts in the same direction.

\section{Measured peak width of plasmon losses}
So far, we have used only the information of peak positions, but we did not discuss the line widths of 
the plasmon losses. These loss peaks turned out to be fairly broad and not limited by the energetic 
resolution of the measurements. We discuss this problem using the recent example of Au/Si(553).

\begin{figure}[tb]
\begin{center}
\includegraphics[width=0.8\textwidth]{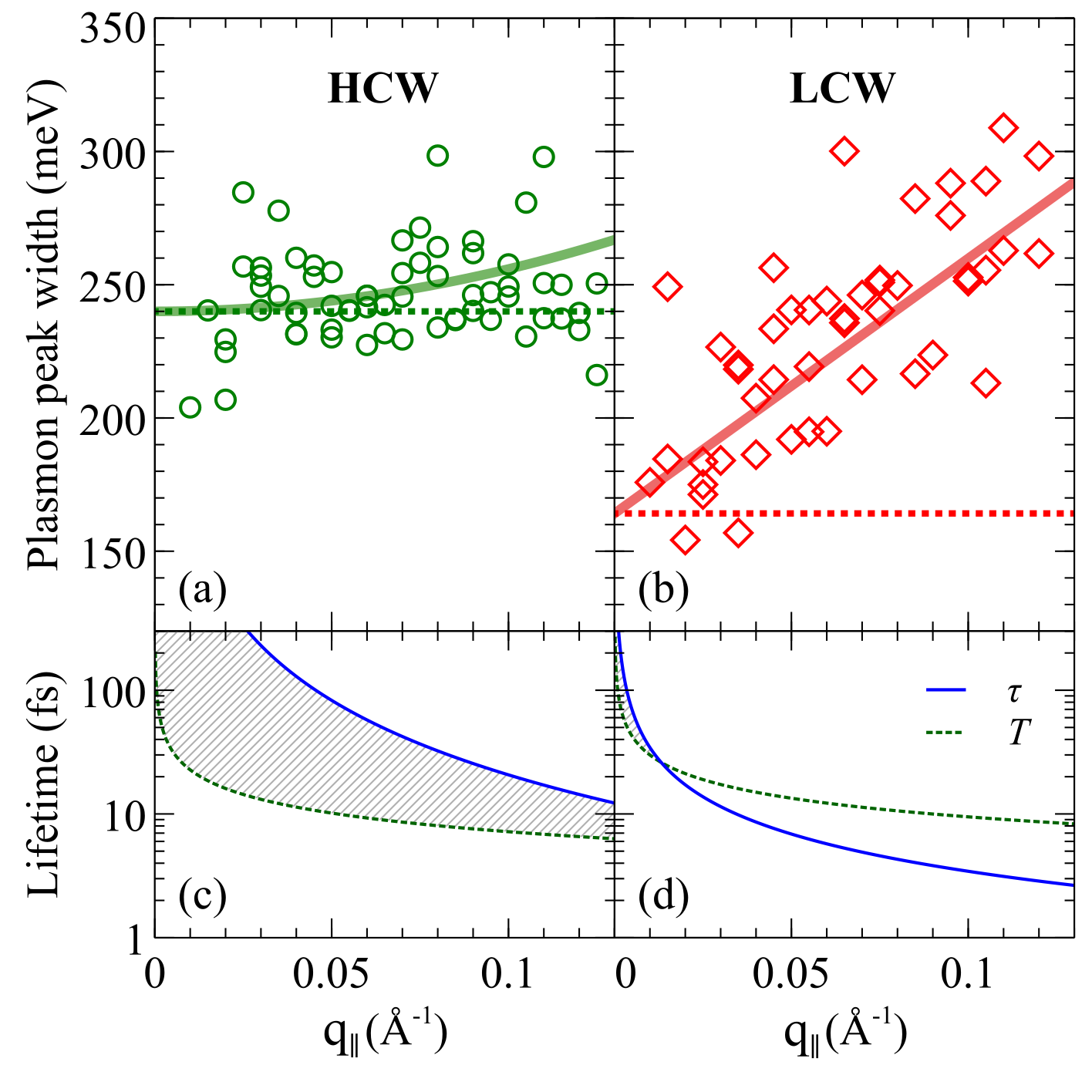}
\end{center}
\caption{Top: Plasmon peak widths of the two systems for varying $q_\parallel$ as 
a measure for the change of plasmon lifetime for the HCW (a) and LCW phase (b).
Bottom: Estimated lifetimes $\tau$ from the data above after the subtraction of the 
dotted offset in comparison with the oscillation period $T$ derived from the dispersion relations 
(see text). \label{fig.width}}
\end{figure}

From  a quick look at the measured width of the loss peaks, as shown in 
Fig.\,\ref{fig.width} for both Au phases on Si(553),  particularly in the  
limit of $q_\parallel \rightarrow 0$, it is already obvious that 
these widths cannot be attributed to a lifetime, since in this limit the plasmonic excitation 
energies also go to zero. Consequently, the life times have to go to infinity. Taking the measured 
width as lifetime would mean that the plasmonic excitation is highly overdamped and plasmonic 
excitations would not exist. This is in sharp contrast to plasmons (at 125 meV) in comparable 
systems, where mean free paths exceeding 200\,nm were observed \cite{Hotzel2015}. 
Therefore, the FWHM at $k=0$ 
must be the convolution of instrumental resolution in reciprocal space and in energy of 
the spectrometer, possibly also with a contribution caused by 
the short interaction time of the electron scattering process. 
The energy resolution of the instrument, however, cannot be the main source of the
measured half widths -- it accounts to only up to 10\% of the measured FWHM. 
Geometrical imperfections, as seen by the finite width in $k$ for the elastic peak, is another 
source, since it will  cause corresponding averaging over the dispersion and thus energetic 
broadening. From systematic studies of LEED profile analysis it can also be safely ruled out here as 
a main source of broadening. 
The finite k-space resolution  of $\pm 0.01$\,\AA$^{-1}$, however, can be the main source of 
broadening, as seen by a comparison with the slope of dispersion of the HCW phase from 
Fig.\,\ref{disp-2}. Since this slope is generally lower by about 30\% for the LCW phase compared 
with the HCW phase, the broading in the limit of $q_\parallel \rightarrow 0$ is correspondingly 
reduced, as indeed seen in Fig.\,\ref{fig.width}. Therefore, the short interaction time of the 
inelastically scattering electrons as 
the possible reason for broadening seems to play a minor role. It may, however, not be negligible 
in the present case, since only near-field multipole scattering can be effective 
here -- dipole scattering cannot transfer the momentum necessary for excitation -- the scattering 
process being closer to impact scattering \cite{Ibach1982}. 

Starting from the FWHM at $q= 0$ as an ``instrumental'' baseline, there is a tendency for a slight 
increase of FWHM by about 30\,meV at $q_\parallel = 0.13$\AA\ for the HCW phase.
This increase is much higher for the LCW phase and amounts to about 120\,meV for the same range of 
$q_\parallel$. Two interpretations  seem to be possible: broadening due to reduced lifetime with 
increasing energy of excitation, or inhomogeneous broadening caused by the existence of two plasmons.
 
For the HCW phase lifetime broadening alone seems to be a possible explanation 
for the observed increase in FWHM as a function of $q_\parallel$, depicted in 
figure~\ref{fig.width}(c). Using the procedures as 
just described, we approximated the increase of the peak widths by polynomial functions of second 
order (solid curves), subtracted the offset (dotted lines), and estimated the lifetime with 
Heisenberg's uncertainty principle of $\Delta E \Delta t \geq \hbar / 2$, which yields the blue 
curve in the lower left part (c) of figure~\ref{fig.width}. This curve represents still a lower 
limit, since no broadening due to slight energetic differences in the two plasmon modes, especially 
at higher $q_\parallel$,  was assumed to exist also here. In any case, it is compatible 
with other data \cite{Hotzel2015}.   

For the LCW phase, on the other hand, lifetime broadening as the main source for the increase in 
FWHM can be ruled out right away. Assuming that we can simply subtract 
the halfwidths of the baseline from those measured,
the determined ``lifetime'' $\tau$ as a function of $q_\parallel$ quickly falls below the time $T$ 
necessary for a single plasmon oscillation, as shown in the lower right part (d)
of figure~\ref{fig.width}. Of course, this overdamped situation is unphysical. Therefore, it seems 
to be more realistic that a large part of the broadening seen is due to the existence of two 
plasmons. 

From the large linewidths of the plasmonic losses observed for both phases and their weak 
dependence on $q_\parallel$ we conclude that the main contribution is not due to short 
life times, but even with the high momentum resolution of the present instrument is still mainly 
caused by the finite instrumental q-resolution, i.e there is still a need for further 
experimental improvement.  While the two plasmons originating from the two Au-induced electronic 
surface bands cannot be resolved in both HCW and LCW phases, their larger difference in excitation 
energy in the LCW phase leads to an additional contribution to the measured halfwidths. 

\section{Conclusions}
Considerable knowledge has been gathered on the dispersion of 2D sheet plasmons over the last 20 years. 
Plasmon dispersion in these systems - a few characteristic examples have been shown above - can surprisingly 
well be described by a nearly-free electron gas, even with the modifications of massless fermions in a 
relativistic electron gas. The inclusion of correlation effects beyond RPA seems to have mainly a
quantitative effect on the general slope of dispersion. At long wavelengths, the dependence on electron density of plasmons 
in a non-relativistic electron gas, on the other hand, seems to follow the 
$\sqrt{n_e}$ dependence in all dimensions. Only in the relativistic case it is reduced to $\propto n^{1/4}$. 

Furthermore,  two main possibilities for linearization 
of plasmon dispersion have been identified that are quite important for potential signal transmission via 
plasmonic coupling. The first one is due to the general form of plasmon dispersion (see eq.\,\ref{eq1}), if the 
second term of this equation dominates. This condition depends on $k_F$, and it can be easily  achieved  
at low electron concentrations and large $q_\parallel$. In the special case of a relativistic electron gas like 
in graphene, it leads to the observed insensitivity of plasmon dispersion on the doping level at large $q_\parallel$. 
In other words, tunability of plasmons in graphene by changes of the doping level only exists at 
long wavelengths (terahertz to far infrared range). The short wavelengths can still be used, but the strategy must 
be the selection of certain wavelengths, which are then quite robust against distortion.  
Particularly on metal surfaces, this range is even more limited due to the large value of the real part of the dielectric 
function of metals at terahertz frequencies.   

Linearization of plasmon dispersion already in the long wavelength limit can be achieved by coupling of 
a 2D electron gas to other 2D or 3D electron gases, leading to the acoustic surface plasmon (ASP). 
This property has been demonstrated for the Shockley surface states on Be(0001) and on the (111) surfaces of noble metals, 
but this concept is fairly general. E.g. it might also be applicable for stacks of 2D metallic sheets separated by insulating
material, or for stacks of graphene sheets, thereby combining linear dispersion with tunability.  

In the anisotropic 2D case, which as a limit includes the quasi-1D wires, the situation less clear. 
While it is obvious that restriction of an electron gas to 1D leads to linearization of the plasmon dispersion already 
to lowest order in $q_\parallel$, there is an explicit dependence of dispersion on the form of the confining potential. 
Furthermore, the crossover into the second dimension by explicit consideration of the wire width and the role of coupling 
of plasmonic excitations with the environment is a general problem the description of which is still imcomplete. 
Also the role of correlations exceeding NFEG properties is still an open question. 
This sensitivity to the embedding environment, on the other hand, opens many still largely unexplored possibilities for 
tuning and manipulation both on the experimental and the theoretical side. 

The present experimental setups, mainly electron loss spectroscopy (HREELS, EELS-LEED) still only allow 
limited studies of dynamics of 2D and 1D plasmons. The main limit is k-space resolution. It is obvious that there is 
sensitivity to atomic-size defects such as steps, which increases as a function of $q_\parallel$ 
\cite{Langer2010}, i.e. there is high sensitivity to the quality of long-range order and its imperfection 
on the atomic scale. 

The questions related to electronic effects in plasmon dynamics, on the other hand, is an 
urgent problem that has been touched qualitatively at best, as our examples show. Since the low-D plasmons are
more strongly damped than, e.g., surface plasmons due to the conductive properties of the metallic sheets and wires, 
but also of their environment, it is clear that solving these questions will be a key issue for any 
application of these plasmons.\\[2ex]

{\noindent\large \bf Acknowledgements}\\
It is a pleasure to acknowledge the experimental contributions of the groups in Hannover and in Genova to several parts of this work, 
in particular by T. Block, U. Krieg, T. Lichtenstein. We also strongly benefitted from collaborations and discussions with 
the theory group at DIPC, San Sebastian, Spain, in particular by the collaboration with V.M. Silkin. 
Financial support to this work by the Deutsche Forschungsgemeinschaft 
mainly through FOR1700 and project Pf238/28   is gratefully acknowledged.

\end{document}